\documentclass[reprint,aps,pre]{revtex4-1}
\usepackage{graphicx}

\usepackage [novolumeabbr] {unitsdef} % The definition of picolitre (\pl) is incompatible with the revtex4 package \pl definition as "Phys. Lett."
\DeclareMathOperator{\cc}{c.c.} % Complex conjugate
\DeclareMathOperator{\NST}{NST} % Nonsecular term
\DeclareMathOperator{\Real}{Re}   % Real part
\newunit{\torr}{Torr}	% Torr - unit of pressure
\newunit{\nanovolt}{nV} % nanovolts
\newunit{\microvolt}{\mu V} % microvolts
\newunit{\Kelvin}{\degree\kelvin} % degrees of Kelvin
\newunit{\gigapascal}{G\pascal} % Pascal * 10^9
\newunit{\megapascal}{M\pascal} % Pascal * 10^6
\newunit{\ppm}{ppm} % Parts per million
\newunit{\nanowatt}{n\watt} % Watt *10^{-9}

\begin{document}
\title {Nonlinear dynamics of a microelectromechanical mirror in an optical resonance cavity}%\\ \emph{Rev. 5}}

\author{Stav~Zaitsev}
\email{e-mail: zzz@tx.technion.ac.il}
% \author{Oleg~Shtempluck}
\author{Eyal~Buks}
\affiliation{Department of Electrical Engineering, Technion - Israel Institute of Technology, Haifa 32000, Israel}
\author{Oded Gottlieb}
\affiliation{Department of Mechanical Engineering, Technion - Israel Institute of Technology, Haifa 32000, Israel}

\begin{abstract}
The nonlinear dynamical behavior of a micromechanical resonator acting as one of the mirrors in an optical resonance cavity is investigated. The mechanical motion is coupled to the optical power circulating inside the cavity both directly through the radiation pressure and indirectly through heating that gives rise to a frequency shift in the mechanical resonance and to thermal deformation. The energy stored in the optical cavity is assumed to follow the mirror displacement without any lag. In contrast, a finite thermal relaxation rate introduces retardation effects into the mechanical equation of motion through temperature dependent terms. Using a combined harmonic balance and averaging technique, slow envelope evolution equations are derived. In the limit of small mechanical vibrations, the micromechanical system can be described as a nonlinear Duffing-like oscillator. Coupling to the optical cavity is shown to introduce corrections to the linear dissipation, the nonlinear dissipation and the nonlinear elastic constants of the micromechanical mirror. The magnitude and the sign of these corrections depend on the exact position of the mirror and on the optical power incident on the cavity. In particular, the effective linear dissipation can become negative, causing self-excited mechanical oscillations to occur as a result of either a subcritical or supercritical Hopf bifurcation. The full slow envelope evolution equations are used to derive the amplitudes and the corresponding oscillation frequencies of different limit cycles, and the bifurcation behavior is analyzed in detail. Finally, the theoretical results are compared to numerical simulations using realistic values of various physical parameters, showing a very good correspondence.
\end{abstract}
\maketitle

% \tableofcontents
% \newpage

%%%%%%%%%%%%%%%%%%%%%%%%%%%%%%%%%%%%%%%%%%%%%%%%%%%%%%
\section{Introduction}
\label{sec:introduction}

The experimental study of interactions between light and mechanical systems was pioneered more than a hundred years ago by Crookes \cite{Crookes_1875}, Lebedew \cite{Lebedew_1901} and others \cite{Nichols&Hull_1901}. The two main coupling mechanisms between radiation and mechanical systems, namely, radiation pressure and thermal effects, were already present in these first experiments. Since then, the effects of radiation pressure have attracted a significant interest.  An early example is the proposition to use the radiation pressure as a driving force in space \cite{Hollerman_02}. Another example comes from the efforts to detect gravitational waves. The optomechanical coupling as a source of additional noise in gravitational waves detectors and the possibility to utilize a high-finesse optomechanical cavity for noise reduction in these detectors has been actively discussed for several decades (see Refs.~\cite{Braginsky&Manukin_67, Braginsky_et_al_70, Kimble_et_al_01, Corbitt_et_al_06, Kippenberg&Vahala_08} and references therein). More recently, similar mechanical mode cooling techniques based on radiation pressure have been proposed as a possible way to quench the thermal noise in a single mechanical vibration mode down to the quantum limit \cite{Schliesser_et_al_08, Genes_et_al_08, Kippenberg&Vahala_08, Teufel_et_al_10}.

The renormalization of the effective mechanical damping due to coupling of a mechanical oscillator to an optical resonance cavity is at the heart of these "cooling" methods. The root cause of the changes in the effective mechanical dissipation in optomechanical systems is the retardation in the radiation induced forces. In many studies, a retardation which occurs in the radiation pressure in optomechanical cavities with high finesse \cite{Braginsky&Manukin_67, Carmon_et_al_05, Arcizet_et_al_06, Gigan_et_al_06, Paternostro_et_al_06, Kippenberg&Vahala_08, Jayich_et_al_08, Marino&Marin_10} is considered.  In such cavities, the optical relaxation rate is comparable to the period of the mechanical oscillations. However, high finesse cavities require state of the art manufacturing technology and, in general, are not readily adjustable for a wide range of different mechanical mirrors. On the other hand, optically induced thermal effects have been shown experimentally to affect the dynamics of optomechanical systems, including  those with finesse of order of unity \cite{Stokes_et_al_90, Hane&Suzuki_96, Metzger&Karral_04, Jourdan_et_al_08,  Metzger_et_al_08, Marino&Marin_10}. In these cases, the retardation is due to a finite thermal relaxation rate \cite{Ludwig_et_al_07, Restrepo_et_al_10, Liberato_et_al_10}.

In contrast with the thoroughly investigated mechanical mode cooling effect, other dynamical phenomena that arise from the optomechanical coupling have received limited theoretical attention. These phenomena include self-excited oscillations  \cite{Hane&Suzuki_96, Braginsky_et_al_01,  Kim&Lee_02, Aubin_et_al_04, Kippenberg_et_al_05, Marquardt_et_al_06, Arcizet_et_al_06, Ludwig_et_al_07, Metzger_et_al_08}, and changes in the effective nonlinear elastic and dissipative behavior of an optomechanical mirror.

As the field of nano optoelectromechanical systems (NOEMS) \cite{Lyshevski&Lyshevski_03, Wu_et_al_06, Hossein-Zadeh&Vahala_10} grows and matures, and, in parallel, the search for mechanical systems at quantum limit intensifies, an increasing number of different optomechanical systems are being investigated. A theoretical model that accurately describes all the phenomena in an optomechanical system and which is able to reproduce the transitional dynamics as well as the steady state and the small vibrations behavior would be of great benefit, especially for the design of such systems and the experimental identification of their parameters.

In this work, we develop a theoretical model of a micromechanical mirror acting as a part of an optical resonance cavity. The mirror is described as a nonlinear oscillator, with cubic elastic and dissipative terms in its equation of motion \cite{Dykman&Krivoglaz_84, Nayfeh_Mook_book_95, Aubin_et_al_04}. The forces acting on the mirror include direct radiation pressure, a thermal force proportional to the temperature change of the mirror, and an external excitation. In addition, a linear dependence of the mechanical resonance frequency on the temperature is assumed. Using a combined harmonic balance and averaging method \cite{Szemplinska-Stupnicka_book_90} to solve the weakly nonlinear equations of motion, we find a practical approximation of this model in the form of evolution equations that describe the slow envelope dynamics of the system.  We investigate two important limiting cases of these general evolution equations.

First, we derive the evolution equations for the case of small vibrations. In addition to the renormalization of the linear mechanical dissipation, we find that the coupling to an optical resonance cavity introduces additional elastic and dissipative nonlinearities into the dynamics of the micromechanical mirror. Based on these results, stability criteria are derived for small oscillations of the mirror, and are shown to coincide with the predictions of a local stability analysis of the full dynamical system. In addition, the small limit cycle amplitude and frequency is given for cases in which a supercritical Hopf bifurcation occurs, and the divergence time scale is estimated for a stability loss process that leads to a subcritical Hopf bifurcation and, consequently, to a jump to a large amplitude limit cycle.

Next, we explore the behavior of the system at finite amplitudes without external excitation. Using the full slow envelope evolution equations, we derive the expressions governing the amplitudes and frequencies of all limit cycles \cite{Strogatz_book_94} that exist in the system. The resulting steady state amplitude equations have the same form as those derived in literature from general power or force balance considerations \cite{Ludwig_et_al_08, Metzger_et_al_08, Lazarus_et_al_10, Aubin_et_al_04}. However, in this work, we are able to formulate the full evolution equations. Therefore, the dynamics of the system can be traced in time, in addition to the final steady state solutions similar to those previously given in the literature.

Finally, we explore the validity of our combined harmonic balance - averaging method and other assumptions. We find that the method is applicable to a wide range of practical optomechanical cavities, especially those in which the finesse is relatively low, the mechanical quality factor is large and the dependence of the mechanical frequency on radiation heating is relatively weak. In contrast, the amplitude of the mechanical mirror vibration does not have to be small, and can be comparable to the optical wavelength or larger. These assumptions are correct for most optomechanical resonators, except for those designed specifically to be incorporated in high finesse optical cavities. However, the mathematical method described here can be readily applied to these systems as well.

In order to experimentally validate the theoretical results derived in this article, we have recently studied an optomechanical cavity with a moving mirror in the form of a freely suspended micromechanical resonator. Using the theoretical model developed here, we have been able to quantitatively describe the dynamics of micromechanical mirrors with two different geometries and material compositions \cite{Zaitsev_et_al_11b}. The theory and the experiment have been found to be in a good agreement both in the domain of forced oscillations and self excitation.

%%%%%%%%%%%%%%%%%%%%%%%%%%%%%%%%%%%%%%%%%%%%%%%%%%%%%%
\section{Theoretical model}

%-------------------------------------------------------------------------------------------------------------------------------
	\subsection{Optomechanical resonance cavity}
	\label{sec:optical_resonance_cavity}

\begin{figure}
	\centering
	 \includegraphics[width=3.4in]{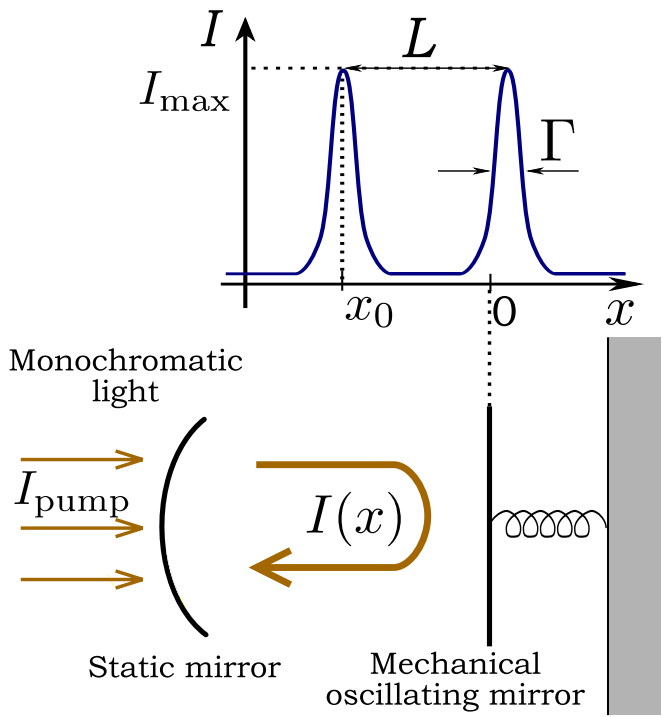}
        \caption{(Color online) A general optomechanical resonance cavity. The left mirror is static. The right mirror is a mechanical oscillator which can move in the direction parallel to the cavity axis ($x$ direction). The cavity is pumped by a constant monochromatic light beam with the power $I_\text{pump}$. The optical power circulating inside the cavity $I$ depends on the actual position of the mechanical mirror, i.e., $I=I(x)$. When the mechanical mirror is at rest, and no light is present, the mirror's position is denoted as $x=0$. The position of the mirror at which the optical power inside the cavity is maximal is called the spatial detuning and is denoted as $x_0$.}
        \label{fig:optomech_cavity}
\end{figure}

Consider an optical resonance cavity constantly pumped by monochromatic laser light, in which one of the mirrors acts as a nonlinear mechanical oscillator (see Fig.~\ref{fig:optomech_cavity}) whose displacement is denoted by $x$. In addition, the cavity medium is considered to be lossless, e.g., vacuum, and all optical losses (such as absorption and diffraction losses) occur at the mirrors.

We refer the reader to the extensive body of literature which exists for an in depth treatment of optical resonance cavities (see, for example, Refs.~\cite{Yariv_book_91, Jackson_book_99, McVoy_et_al_67, Sumetsky&Eggleton_03, Viviescas&Hackenbroich_03} and references therein). Here, we state the results which are needed in order to describe a simple optomechanical system.

If the energy stored in the optical cavity in steady state reaches a local maximum at $x=x_0$, the intra-cavity optical power incident on a mirror can be written as~\cite{Yariv_book_91}
\begin{equation}
	I(x)=\frac{I_\text{max}\left(\frac{\Gamma}{2}\right)^2}{\frac{L^2}{2\pi^2}\left[1-\cos 2\pi\frac{x-x_0}{L}\right]+\left(\frac{\Gamma}{2}\right)^2},
	\label{eq:opt_power_periodic_Lorentzian}
\end{equation}
where $\Gamma$ is the full width at half maximum parameter, $L$ is the distance between two successive resonant positions of the micromechanical mirror, and
\begin{equation*}
	I_\text{max}=C_\text{re}I_\text{pump},
\end{equation*}
where $I_\text{pump}$ is the power of the monochromatic light incident on the cavity, and $C_\text{re}$ is the ratio of the resonant enhancement of the intra-cavity power. Note that for an empty cavity with metallic mirrors,
\begin{equation}
	L=\lambda/2,
	\label{eq:large_osc_L}
\end{equation}
where $\lambda$ is the optical wavelength. In addition, the finesse of the optical cavity can be expressed as 
\begin{equation*}
	\mathcal F =\frac{L}{\Gamma}.
	\label{eq:finesse}
\end{equation*}

If the maximum mechanical displacement $\max{|x|}$ is significantly smaller than $\Gamma$, a quadratic approximation for $I(x)$ can be employed. In this case,
\begin{subequations}
\begin{equation}
	I_0 = I(x=0)=\frac{I_\text{max}\left(\frac{\Gamma}{2}\right)^2}{\frac{L^2}{2\pi^2}\left[1-\cos 2\pi\frac{x_0}{L}\right]+\left(\frac{\Gamma}{2}\right)^2},
% 	I_0 = I(x=0) =\frac{hI_\text{max}}{1+h-\cos 2\pi\frac{x_0}{L}},
\end{equation}
and
\begin{equation}
	I(x) \approx I_0+I_0'x+\frac12 I_0''x^2.
\end{equation}
where a prime denotes differentiation with respect to the displacement $x$.
% \begin{align*}
% 	I_0' & = \frac{2\pi}{L}\frac{I_0 \sin\left(2\pi\frac{x_0}{L}\right)}{1+h-\cos \left(2\pi\frac{x_0}{L}\right)},\\
% 	I_0'' & = \frac{2\pi}{L}\frac{\frac{2\pi}{L}I_0 \cos\left(2\pi\frac{x_0}{L}\right)+I_0' \sin\left(2\pi\frac{x_0}{L}\right)}{1+h-\cos \left(2\pi\frac{x_0}{L}\right)}.
% \end{align*}
	\label{eq:I_approx_parameters}
\end{subequations}

In this work, the optical power is assumed to follow the displacement without any lag. Namely, the optical response time is assumed to be much shorter than any other timescale in the system, including thermal relaxation time and mechanical vibration period.
% Such an assumption is justified, for example, for an infrared resonance cavity in which the finesse is of order of $10^4$ (or lower), the optical cavity length is tens of wavelengths, and the mechanical resonance frequency of the micromechanical mirror is in the megahertz range.

The function $I(x)$ in Eq.~\eqref{eq:opt_power_periodic_Lorentzian} can be represented by spatial Fourier series,
\begin{equation}
	I(x)=\sum_{k=-\infty}^\infty c_k e^{j2\pi k\frac{x}{L}},
	\label{eq:large_osc_I_Fourier}
\end{equation}
where
\begin{equation}
	c_k=\frac{1}{L}\int_0^{L} I(x)e^{-j2\pi k\frac{x}{L}}dx.
	\label{eq:large_osc_ck}
\end{equation}
Note that $c_k=c_{-k}^*$ because $I(x)$ is real.

In practice, if the optical power $I(x)$ changes relatively slowly with displacement, the series in Eq.~\eqref{eq:large_osc_I_Fourier} converges quickly (an example is shown in Fig,~\ref{fig:large_osc_I_series}), and Eq.~\eqref{eq:large_osc_I_Fourier} can be approximated as
\begin{equation}
	I(x) \approx \sum_{k=-k_\text{max}}^{k_\text{max}} c_k e^{j2\pi k\frac{x}{L}},
	\label{eq:large_osc_I_Fourier_kmax}
\end{equation}
where $k_\text{max} \gtrsim \mathcal F$. The exact expressions for $c_k$ are derived in Appendix~\ref{app:opt_power_Fourier_series}.

%There, it is also shown that the truncation error quickly becomes negligible for increasing $k_\text{max}$ for relatively wide ($\Gamma>0.01L$) optical resonances (see Fig.~\ref{fig:large_osc_I_series_error}).

%-------------------------------------------------------------------------------------------------------------------------------
	\subsection{Equations of motion}
	\label{sec:eq_of_motion}
We model the dynamics of the micromechanical mirror in the optical cavity by approximating it by a nonlinear mechanical oscillator with a single degree of freedom $x$ operating near its primary resonance~\cite{Zaitsev_et_al_11}. The mechanical oscillator's equation of motion is given by
\begin{multline}
	\ddot x+\frac{\omega_0}{Q}\dot x+\omega_m^2x+\alpha_3 x^3+\gamma_3 x^2\dot x \\
	=2f_m\cos(\omega_0+\sigma_0)t+F_\text{rp}(x)+F_\text{th}(x),
	\label{eq:mech_osc_orig}
\end{multline}
where a dot denotes differentiation with respect to time $t$, $x$ is the mirror displacement, $\omega_0$ is the original resonant frequency of the mirror, $Q$ is the mechanical quality factor, $\omega_m$ is the momentary resonance frequency, whose dependence on $\omega_0$ and other parameters will be discussed below, $\alpha_3$ is the nonlinear (cubic) elastic coefficient, and $\gamma_3$ is the nonlinear dissipation coefficient. In addition, $f_m$ is the external excitation force, $\sigma_0$ is a small detuning of the external excitation frequency from $\omega_0$, $F_\text{rp}$ is a force resulting from radiation pressure, and $F_\text{th}$ is a force resulting directly from temperature changes in the micromechanical mirror (such force can be attributed, for example, to thermal deformations \cite{Metzger_et_al_08} or buckling).

Below, we consider external excitation frequency detuning $\sigma_0$ to be small, i.e., $\sigma_0\ll\omega_0$. In addition, the mechanical quality factor is assumed to be large, i.e., $Q\gg1$.

It has been shown previously that nonlinear effects can play an important role in the dynamics of micromechanical systems  \cite{Lifshitz&Cross_08, Zaitsev_et_al_11}. In our case, we assume that the micromechanical mirror behaves as a Duffing-like oscillator with positive nonlinear dissipation $\gamma_3>0$ (i.e., the uncoupled autonomous mechanical system ($f_m=0$) is unconditionally stable). Note that throughout this study, the mechanical nonlinearities are assumed to be weak, i.e., $\alpha_3 x^2 \ll \omega_m^2$ and $\gamma_3 x^2 \ll \omega_0/Q$.

% The optical power inside the cavity is a function of the displacement $x$. We denote this power by $I(x)$. The exact expression for $I(x)$ in form of a periodic Lorentzian-like function was given in Sec.~\ref{sec:optical_resonance_cavity}, but the analysis below can be applied to any reasonable optical power function which is periodic in the displacement $x$.

We assume linear dependence of the mechanical resonance frequency on the temperature:
\begin{equation}
	\omega_m=\omega_0-\beta(T-T_0),
	\label{eq:omega_m}
\end{equation}
where $\beta$ is a proportionality coefficient, $T$ is the effective temperature of the mechanical oscillator, and $T_0$ is the temperature of the environment. In the majority of experimental situations, $\beta$ is positive, i.e., heating of the micromechanical oscillator reduces its resonance frequency, while cooling increases it.

In general, the nonlinear coefficients $\alpha_3$ and $\gamma_3$ are functions of temperature similarly to $\omega_m$. However, due to the fact that the nonlinear terms are assumed to be small in Eq.~\eqref{eq:mech_osc_orig} and the impact of their thermal variation is much smaller than that of $\omega_m$, we regard the nonlinear mechanical coefficients as constants. The same is true for the linear dissipation coefficient $\omega_0/Q$.

The time evolution of the effective temperature is governed by the following equation:
\begin{equation}
	\dot T=\kappa(T_0-T)+\eta I(x),
	\label{eq:temp_diff_eq}
\end{equation}
where
\begin{equation}
	\eta=\frac{h_\text{rad}}{mC_m},
\end{equation}
the effective mass of the oscillator is denoted by $m$, $h_\text{rad}$ is the radiation absorption factor of the mirror material, $C_m$ is the mass-specific heat capacity of this material, $\eta$ is the heating rate due to interaction between the material of the mechanical oscillator and the light in the optical cavity, $T_0$ is the temperature of the environment and $\kappa$ is the effective thermal conductance coefficient.
%, and a dot denotes differentiation with respect to time.
In this simple approximation, the nonuniform temperature distribution due to localized radiative heating in the micromechanical mirror is disregarded.

In general, in addition to radiative heating term $\eta I(x)$, Eq.~\eqref{eq:temp_diff_eq} should account for heating due to mechanical damping. The heating power of this process can be estimated as
\begin{equation*}
	P_Q \approx \frac{1}{\tau}\frac{m\omega_0^2A^2}{2Q},
\end{equation*}
where $\tau=2\pi/\omega_0$ is the time period of the mechanical vibrations, $A$ is the amplitude of these vibrations, and all nonlinear effects have been neglected for simplicity. Comparing this heating power to the heating term in Eq.~\eqref{eq:temp_diff_eq}, we find that $P_Q$ is generally negligible if
\begin{equation}
	\frac{P_Q}{h_\text{rad}I(x)}=\frac{\frac{m\omega_0^2A^2}{2Q}}{\tau h_\text{rad}I(x)} \ll 1.
\end{equation}
For example, for typical values of $\omega_0=10^6\sec^{-1}$, $m=10^{-11}\kg$, $Q=10^5$, and $A=1\micm$, we find that $P_Q\approx 8\times 10^{-12}\watt$. We compare this to the radiative heating by assuming that the radiation absorption factor $h_\text{rad}$ of the micromechanical mirror is of order of several percents. It follows that if the optical power $I$ in the cavity is approximately $10\nanowatt$ or higher, the radiative heating is the dominant heating process. In practice, the optical powers that can have a significant impact on the system's dynamics and that are used in the experiments are of order of microwatts or higher, and, therefore, a term proportional to $P_Q$ is neglected in Eq.~\eqref{eq:temp_diff_eq}.

The formal solution of Eq.~\eqref{eq:temp_diff_eq} is
\begin{equation}
	T(t)=e^{-\kappa t}\left[\int_0^t \left(\kappa T_0+\eta I(x)\right) e^{\kappa\tau}d\tau+T(t=0)\right].
\end{equation}
This can be shown to result in
\begin{equation}
	T-T_0=\eta\int_0^t I(x) e^{\kappa(\tau-t)}d\tau,
	\label{eq:T-T0_orig}
\end{equation}
where the initial transient response term $e^{-\kappa t}\left[T(t=0)-T_0\right]$ has been dropped as insignificant to the long timescale dynamics of the system.
% where it has been assumed that $T(t=0)=T_0$, i.e., the initial temperature of the mechanical oscillator is equal to the temperature of the environment.

Using the fact that the energy and the momentum of a photon follow the relation $\mathcal{E}_\text{photon}=cp_\text{photon}$, where $c$ is the velocity of light, we find that the radiation pressure force is
\begin{equation}
	 F_\text{rp}(x)=\nu I(x),
	\label{eq:F_rp} \\
\end{equation}
where
\begin{equation*}
	\nu =\frac{2}{mc},
\end{equation*}
and where 
%$m$ is the effective mass of the mechanical oscillator and
light absorption by the micromechanical mirror has been neglected.

Finally, we introduce a temperature dependent force, which acts directly on the micromechanical mirror. In practice, this thermal force can arise from several effects, such as a deflection of a bimorph mirror layer due to heating, or a distortion due to internal stress~\cite{Guckel_et_al_92, Fang&Wickert_94} caused by a non uniform heating of the mirror. The thermal force $F_\text{th}$ is assumed to be linear in the temperature difference $T-T_0$, i.e.,
\begin{equation}
	F_\text{th}=\theta(T-T_0)=\theta\eta\int_0^t I(x) e^{\kappa(\tau-t)}d\tau,
	\label{eq:Fth}
\end{equation}
where Eq.~\eqref{eq:T-T0_orig} has been used.

The equation of motion~\eqref{eq:mech_osc_orig} can rewritten in a closed form as
\begin{multline}
	\ddot x+\frac{\omega_0}{Q}\dot x+\left[\omega_0-\beta\eta K(I)\right]^2x+\alpha_3x^3+\gamma_3 x^2\dot x \\
	=2f_m\cos(\omega_0+\sigma_0)t+\nu I(x)+\theta\eta K(I),
	\label{eq:full_eq_of_motion}
\end{multline}
where we have defined the functional
\begin{equation}
	K(f) \equiv \int_0^t f(t) e^{\kappa(\tau-t)}d\tau.
	\label{eq:K_memory_kernel}
\end{equation}

Before application of the combined harmonic balance - averaging method to Eq.~\eqref{eq:full_eq_of_motion}, we conduct a stability analysis of the full dynamical system defined by Eqs.~\eqref{eq:mech_osc_orig} and \eqref{eq:temp_diff_eq} in App.~\ref{app:equilibrium_eqs_motion}. There, it is shown that Hopf bifurcation is possible in the original system, and the necessary and sufficient conditions for this bifurcation are derived. These conditions will be shown below to be very similar to those found using the slow varying evolution equations.

%---------------------------------------------------------------------------------------------------------------------------------
	\subsection{High thermal conduction limit}
	\label{sec:large_kappa}
For the case where the characteristic thermal relaxation time $\kappa^{-1}$ is much smaller than any other time scale in the system, namely $\omega_{0,m}^{-1}$ and $Q/\omega_0$, the equation of motion~\eqref{eq:full_eq_of_motion} can be significantly simplified. The memory kernel $e^{\kappa(\tau-t)}$ in Eq.~\eqref{eq:K_memory_kernel} can be replaced by a delta function $\delta(\tau-t)/\kappa$, i.e.,
\begin{equation}
	T-T_0=\frac{\eta}{\kappa}I(x).
	\label{eq:T-T0_large_kappa}
\end{equation}
Consequently, the equation of motion~\eqref{eq:full_eq_of_motion} becomes
\begin{multline}
	\ddot x+\frac{\omega_0}{Q}\dot x+\left[\omega_0-\frac{\beta\eta}{\kappa}I(x)\right]^2x+\alpha_3x^3+\gamma_3 x^2\dot x\\
	-\left(\nu-\frac{\theta\eta}{\kappa}\right) I(x)=2f_m\cos(\omega_0+\sigma_0)t,
	\label{eq:eq_of_motion_large_kappa}
\end{multline}

It is easy to see that if the thermal relaxation rate in the system is fast compared to the mechanical resonance frequency, then the sole result of the coupling between the mechanical system and the optical cavity is the addition of nonlinear elastic terms proportional to $I(x)$, $I(x)x$ and $I(x)^2x$ in the mechanical equation of motion~\eqref{eq:eq_of_motion_large_kappa}. The mechanical dissipation terms proportional to $\omega_0/Q$ and $\gamma_3 x^2$ remain unchanged.

%---------------------------------------------------------------------------------------------------------------------------------
	\subsection{Finite amplitude oscillations analysis}
	\label{sec:large_osc_theory}
In general, in order for dissipative terms to occur in an equation of motion, some retardation in the displacement dependent force acting on the system is required \cite{Ludwig_et_al_07, Kippenberg&Vahala_08}. In our case, it is the memory kernel integral in $K(I)$ in Eq.~\eqref{eq:full_eq_of_motion} that provides this retardation. In other words, the finite thermal relaxation rate $\kappa$ and the coupling of momentary mechanical resonance frequency $\omega_m$ to the optical power $I(x)$ can be expected to result in changes in the effective linear and nonlinear dissipation of the micromechanical mirror [see Eqs.~\eqref{eq:omega_m}, \eqref{eq:temp_diff_eq}, and \eqref{eq:K_memory_kernel}].
	
It follows from the above discussion that a nontrivial dissipation behavior can be expected when the rate of thermal relaxation $\kappa$ is comparable to the mechanical resonance frequency $\omega_m$.  We investigate the dynamics of mechanical oscillations with arbitrary amplitudes, i.e., oscillations with amplitudes that can be comparable with the wavelength of the light. The behavior of the optical power $I$ as periodic function of the displacement $x$ has been described in Sec.~\ref{sec:optical_resonance_cavity}.

In order to solve the equation of motion~\eqref{eq:full_eq_of_motion}, we make use of a combined harmonic balance - averaging method~\cite{Szemplinska-Stupnicka_book_90}.
%a method very similar the standard Krylov-Bogoliubov averaging technique \cite{Nayfeh_book_81} will be used.

It can be expected that if all the nonlinear and optic related terms in Eq.~\eqref{eq:full_eq_of_motion} are relatively small then the motion of the mirror is very similar to the motion of a simple harmonic oscillator, i.e.,
\begin{equation}
	x(t) \approx A_0+A_1\cos\psi,
	\label{eq:large_osc_x}
\end{equation}
where
\begin{equation}
	\psi=\omega_0 t+{\tilde\phi},
	\label{eq:large_osc_psi}
\end{equation}
and where $A_1$ and $\tilde\phi$ are the oscillator's amplitude and phase, respectively, and $A_0$ is the static displacement. Here, it is assumed that the amplitude $A_1$ and the phase $\tilde\phi$ do not vary significantly on a time scale defined by $\omega_0^{-1}$ and, therefore, can be considered constant during a single period of the mechanical oscillation. This assumption is commonly referred to as the slow envelope approximation.

The details of the averaging process used to derive the slow envelope evolution equations are given in App.~\ref{app:averaging_eqs_motion}. Here, we state the main results.

Assuming all the frequency corrections as well as the static displacement $A_0$ to be small, we find that [see Appendix~\ref{app:averaging_eqs_motion}]
\begin{equation}
	A_0 \approx \frac{1}{\Omega^2+\frac32\alpha_3A_1^2}\left[2P_1\beta\eta \frac{\omega_0\kappa}{\kappa^2+\omega_0^2}A_1+P_0 \left(\nu+\frac{\theta\eta}{\kappa}\right)\right],
	\label{eq:large_osc_A0}
\end{equation}
where
\begin{equation}
	\Omega=\omega_0-\frac{\beta\eta}{\kappa}P_0=\omega_0-\Delta\omega_0,
	\label{eq:large_osc_Omega}
\end{equation}
and
\begin{equation}
	P_n(A_0,A_1)=\sum_{k=-k_\text{max}}^{k_\text{max}} j^n c_k e^{j2\pi k\frac{A_0}{L}} J_n\left(2\pi k\frac{A_1}{L}\right),
	\label{eq:large_osc_Pn}
\end{equation}
where $J_n(z)$ is the Bessel function of order $n$. The term $\Delta\omega_0$ represents a small mechanical frequency correction due to the averaged heating of the micromechanical mirror vibrating with an amplitude $A_1$.

The evolution equations are (see App.~\ref{app:averaging_eqs_motion}):
\begin{subequations}
\begin{multline}
	\dot A_1 = -\left(\frac{\omega_0}{2Q}+\frac{\gamma_3}{2} A_0^2+2P_2\beta\eta\frac{\omega_0}{\kappa^2+4\omega_0^2}\right)A_1\\
-\frac{\gamma_3}{8}A_1^3-P_1\eta\frac{\omega_0}{\kappa^2+\omega_0^2}\left(2\beta A_0+\frac{\theta}{\omega_0}\right)\\
-\frac{f_m}{\omega_0} \sin\phi,
	\label{eq:large_osc_polar_evolution_dotA1}
\end{multline}
and
\begin{multline}
	A_1\dot \phi =-\left(\sigma_0+\Delta\omega_0-\frac{3\alpha_3}{2\omega_0}A_0^2+P_2\beta\eta \frac{\kappa}{\kappa^2+4\omega_0^2}\right)A_1\\
+\frac{3\alpha_3}{8\omega_0}A_1^3-P_1\eta\frac{\kappa}{\kappa^2+\omega_0^2}\left(2\beta A_0+\frac{\theta}{\omega_0}\right)-P_1\frac{\nu}{\omega_0}\\
-\frac{f_m}{\omega_0} \cos\phi,
	\label{eq:large_osc_polar_evolution_phi}
\end{multline}
\label{eq:large_osc_polar_evolution_eq}%
\end{subequations}
where a new slow varying phase variable has been defined as (recall that the detuning $\sigma_0$ is assumed small)
\begin{equation*}
	\phi=\tilde\phi-\sigma_0 t.
\end{equation*}
% Note that in the case of the non excited system ($f_m=0$, $\sigma_0=0$), the two slow varying phases coincide, i.e., $\tilde\phi=\phi$.

Equations~\eqref{eq:large_osc_polar_evolution_eq} together with Eqs.~\eqref{eq:large_osc_Omega} and \eqref{eq:large_osc_A0} constitute a coupled set of first order differential equations describing the time evolution of the slow envelope of the solution of Eq.~\eqref{eq:full_eq_of_motion}. Now, we proceed to explore two important special cases of the system's behavior - the dynamics at small oscillation amplitudes and the steady state solutions corresponding to various limit cycles.

%--------------------------------------------------------------------------------------------------------------------------------
	\subsection{Small amplitude oscillations limit}
	\label{sec:small_osc_theory}

The equations~\eqref{eq:large_osc_A0} and \eqref{eq:large_osc_polar_evolution_eq} can be significantly simplified for small oscillation amplitudes and static deflections, i.e., for $A_{0,1}\ll \Gamma$. To this end, we denote the oscillation amplitude $A_1$ as $A_{1s}$ in this section and simplify Eqs.~\eqref{eq:large_osc_Omega} and~\eqref{eq:large_osc_A0} to [see also Eqs.~\eqref{eq:I_approx_parameters}]:
\begin{subequations}
\begin{align}
	& \Delta\omega_{0s} =\frac{\beta\eta}{\kappa}I_0, \label{eq:small_osc_Dw0s}\\
	& \Omega_s =\omega_0-\Delta\omega_{0s}, \\
% 	& \sigma_{1s} =\Delta\omega_{0s}+\sigma_0, \\
	& A_{0s} = \frac{I_0}{\Omega_s^2}\left(\nu+\frac{\theta\eta}{\kappa}\right), \label{eq:small_osc_A0s}
\end{align}
	\label{eq:small_osc_params}%
\end{subequations}
where the oscillation frequency $\Omega_s$ and the static deflection $A_{0s}$ are independent of the oscillation amplitude $A_{1s}$. 

In this limit, $P_n$ can be represented by the lowest order terms in its Taylor series expansion, i.e.,
\begin{equation*}
	P_n(A_{1s}) \approx \sum_{k=-k_\text{max}}^{k_\text{max}} j^n c_k J_n\left(2\pi k\frac{A_1}{L}\right).
% 	\label{eq:small_osc_Pn}
\end{equation*}
\begin{equation}
	P_n \approx \sum_{k=-k_\text{max}}^{k_\text{max}} \frac{c_k}{n!} A_{1s}^n \left(j\frac{\pi k}{L}\right)^n\left[1-\frac{A_{1s}^2}{n+1}\left(\frac{\pi k}{L}\right)^2\right].
	\label{eq:small_osc_Pn}
\end{equation}
Using the fact that
\begin{equation}
	\frac{d^nI(x)}{dx^n} \approx \sum_{k=-k_\text{max}}^{k_\text{max}}\left(j\frac{2\pi k}{L}\right)^nc_k,
\end{equation}
we can make the following substitutions for $P_{0,1,2}$:
\begin{subequations}
\begin{align}
	P_0 & \approx  \sum_{k=-k_\text{max}}^{k_\text{max}} c_k\left(1-\left(\frac{\pi k}{L}\right)^2 A_{1s}^2\right) \approx I_0
+\frac14 I_0''A_{1s}^2, \\
	P_1 & \approx \sum_{k=-k_\text{max}}^{k_\text{max}} c_k j\frac{\pi k}{L} A_{1s} \approx \frac{1}{2}I_0'A_{1s}, \\
	P_2 & \approx \sum_{k=-k_\text{max}}^{k_\text{max}} c_k \left(j\frac{\pi k}{L}\right)^2 \frac{A_{1s}^2}{2} \approx \frac18 I_0''A_{1s}^2.
\end{align}
\end{subequations}

Consequently, the equations for $\Omega$ and $A_0$ [Eqs.~\eqref{eq:large_osc_Omega} and \eqref{eq:large_osc_A0}, respectively] can be expanded up to the second order in $A_{1s}$ and first order in $\Delta\omega_{0s}$ resulting in
\begin{subequations}
\begin{equation}
	\Delta\omega_0 \approx \Delta\omega_{0s}+\frac{\beta\eta}{4\kappa}I_0''A_{1s}^2,
\end{equation}
and
\begin{equation}
	A_0 \approx A_{0s}+\left[\beta\eta\frac{\omega_0\kappa}{\kappa^2+\omega_0^2}I_0'+\frac{1}{4}\left(\nu+\frac{\theta\eta}{\kappa}\right)I_0''\right]\frac{A_{1s}^2}{\Omega_s^2}.
\end{equation}
\end{subequations}

Finally, Eqs.~\eqref{eq:large_osc_polar_evolution_eq} can be simplified to
\begin{subequations}
\begin{align}
	\dot A_{1s} & =-\gamma A_{1s}-\frac{r}{4}A_{1s}^3-\frac{f_m}{\omega_0} \sin\phi, \\
	A_{1s}\dot\phi & =-\left(\sigma_0+\Delta\omega_s\right)A_{1s}+\frac{q}{4}A_{1s}^3-\frac{f_m}{\omega_0} \cos\phi.
\end{align}
	\label{eq:small_osc_polar_evolution_eq}
\end{subequations}
where
\begin{subequations}
\begin{equation}
	\gamma =\frac{\omega_0}{2Q}+\frac{\gamma_3}{2}A_{0s}^2+\eta\frac{\omega_0}{\kappa^2+\omega_0^2}\left(\beta A_{0s}+\frac{\theta}{2\omega_0}\right)I_0',
	\label{eq:small_osc_gamma}
\end{equation}
\begin{multline}
	\Delta\omega_s =\Delta\omega_{0s}-\frac{3\alpha_3}{2\omega_0}A_{0s}^2\\
	+\left[\frac{\nu}{2\omega_0}+\frac{\eta\kappa}{\kappa^2+\omega_0^2}\left(\beta A_{0s}+\frac{\theta}{2\omega_0}\right)\right]I_0', \label{eq:small_osc_omegas}
\end{multline}
	\label{eq:small_osc_gamma_and_Deltaws}
\end{subequations}
and
\begin{subequations}
\begin{align}
	q & =\frac{3\alpha_3}{2\omega_0}-\frac{\beta\eta}{2\kappa}\frac{3\kappa^2+8\omega_0^2}{\kappa^2+4\omega_0^2}I_0'',
	\label{eq:small_osc_q} \\
	r & =\frac{\gamma_3}{2}+\beta\eta\frac{\omega_0}{\kappa^2+4\omega_0^2}I_0''.
	\label{eq:small_osc_r}
\end{align}
	\label{eq:small_osc_q_and_r}
\end{subequations}

It is customary to rewrite the evolution equations~\eqref{eq:small_osc_polar_evolution_eq} in a complex form by defining the complex amplitude
\begin{subequations}
\begin{align}
	& a_s =\frac12 A_{1s}e^{j\phi},\\
	& \dot a_s =\frac12 \left( \dot A_{1s}+jA_{1s}\dot\phi \right)e^{j\phi},\\
	& A_{1s}\cos\psi =a_se^{j(\omega_0+\sigma_0)t}+\cc,
\end{align}
\end{subequations}
where $\cc$ denotes a complex conjugate. Using these definitions, the complex evolution equation reads:
\begin{equation}
	j\dot a_s+(j\gamma-\sigma_0-\Delta\omega_s)a_s+(q+jr)a_s^2a_s^*=\frac{f_m}{2\omega_0},
	\label{eq:small_osc_complex_evolution_eq}
\end{equation}

% For completeness, we also rewrite the original evolution equations~\eqref{eq:small_osc_polar_evolution_eq} as
% \begin{subequations}
% \begin{align}
% 	\dot A_{1s} & =-\gamma A_{1s}-\frac{r}{4}A_{1s}^3-\frac{f_m}{\omega_0}\sin(\phi-\sigma_0t), \\
% 	A_{1s}\dot\phi & =-\Delta\omega_sA_{1s}+\frac{q}{4}A_{1s}^3-\frac{f_m}{\omega_0}\cos(\phi-\sigma_0t).
% \end{align}
% 	\label{eq:small_osc_polar_evolution_eq_qr}
% \end{subequations}

Evidently, the coupling of a micromechanical mirror to an optical cavity introduces two types of terms into the complex evolution equation~\eqref{eq:small_osc_complex_evolution_eq} - linear terms proportional to $I_0'$ and nonlinear terms of the third order proportional to $I_0''$. In addition, the autonomous part of the complex slowly varying evolution equation ($f_m=0$) consists of an approximated Hopf normal form of the original system~\cite{Guckenheimer&Holmes_book_83, Strogatz_book_94}, and is expected to yield conditions for self-excited limit cycles following either a sub or supercritical bifurcation determined by the sign of the cubic damping coefficient $r$.

%%%%%%%%%%%%%%%%%%%%%%%%%%%%%%%%%%%%%%%%%%%%%%%%%%%%%%%%%%%%
\section{Small oscillations behavior}
\label{sec:small_osc_behavior}
%----------------------------------------------------------------------------------------------------------------------------------------------
	\subsection{Linear and nonlinear effects in the dynamics of the small oscillations}
	\label{sec:small_osc_dynamics}

The linear terms governing the dynamics of the micromechanical mirror considered here are given in Eqs.~\eqref{eq:small_osc_gamma_and_Deltaws}. The parameter $\Delta\omega_s$ describes a small additional resonance frequency correction which arises from changes in heating and elastic nonlinearity due to small static displacement $A_{0s}$. In general, this correction can be considered small, i.e., $\Delta\omega_s\ll \omega_0$. In contrast, the linear dissipation coefficient $\gamma$ can undergo significant changes as function of the optical power, resulting in qualitative changes in the system's dynamics.

An optical power dependent effective quality factor $Q_\text{eff}$ can be defined by
\begin{multline}
	\frac{1}{Q_\text{eff}}=\frac{2\gamma}{\Omega_s}\\
	=\frac{1}{\Omega_s}\left[\frac{\omega_0}{Q}+\gamma_3 A_{0s}^2+2\eta \frac{\omega_0}{\kappa^2+\omega_0^2}\left(\beta A_{0s}+\frac{\theta}{2\Omega_s}\right)I_0'\right].
	\label{eq:small_osc_Qeff}
\end{multline}
Note that from the experimental point of view, the definition of an effective quality factor given above is convenient because $Q_\text{eff}$ can be extracted directly from the small amplitude free ring down measurements of the micromechanical mirror. In addition, $Q_\text{eff}$ is a function of $I_0'$. It follows that the local properties of $I(x)$ in the vicinity of $x=0$ have a profound impact on the effective linear dissipation of the system. If the micromechanical mirror is positioned at the negative slope of the optical response curve, i.e., if $I_0'<0$, and optical power is large enough, then the effective linear dissipation can be significantly reduced, resulting in extremely large ring down times, or even become negative. Alternatively, if the mirror is positioned at the positive slope, i.e. if $I_0'>0$, a significant increase in the effective dissipation, also known as "mechanical mode cooling", can be achieved (see the discussion and references given in the Introduction section of this article).

The possibility of a negative linear damping suggests that the micromechanical mirror can develop self-excited oscillations. This mode of operation will be further investigated in following sections. Here, we calculate the threshold conditions for the linear damping $\gamma$ to become negative, namely, the value $I_\text{th}$ of $I_\text{max}$ and the value $x_\text{0th}$ of $x_0$ at the threshold.

Neglecting all nonlinear terms and terms proportional to $\Delta\omega_0/\omega_0$, the self oscillation threshold condition at an arbitrary value of $x_0$, as can be derived from Eqs.~\eqref{eq:small_osc_gamma} and~\eqref{eq:small_osc_params}, is
\begin{equation}
	\frac{\omega_0}{2Q}+\frac{\eta}{\kappa^2+\omega_0^2}\left(\frac{\beta\nu}{\omega_0}I_0+\frac{\theta}{2}\right)I_0'=0.
	\label{eq:small_osc_self_osc_threshold_cond}
\end{equation}
It should be emphasized that under the assumptions described above, this condition coincides with the exact Hopf criterion in Eq.~\eqref{eq:app_equilibrium_Hopf_cond} found in App.~\ref{app:equilibrium_eqs_motion} for the original dynamical system defined by Eqs.~\eqref{eq:mech_osc_orig} and \eqref{eq:temp_diff_eq}.

\begin{figure}
	\centering
	 \includegraphics[width=3.4in]{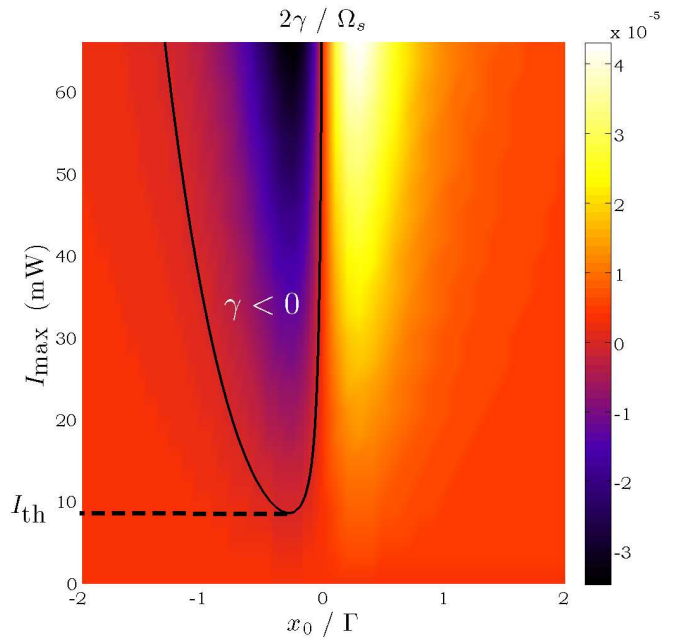}
        \caption{(Color online) Linear dissipation coefficient $\gamma$ vs. the spatial cavity detuning $x_0$ and the optical power $I_\text{max}$. For the positive values of $\gamma$, the non dimensional parameter drawn, $2\gamma/\Omega_s$, is equal to the reciprocal of the effective quality factor $Q_\text{eff}$ [see Eq.~\eqref{eq:small_osc_Qeff}]. In the area above the thick black line, the linear damping is negative ($\gamma<0$), i.e., the solution $x=0$ is no longer stable, Note that for positive values of the spatial detuning $x_0$, the linear damping can greatly exceed the pure mechanical value, $1/Q=0.4\times 10^{-5}$.}
        \label{fig:small_osc_gamma}
\end{figure}

For a system in which the thermal force is dominant, the term proportional to $\nu$ in Eq.~\eqref{eq:small_osc_self_osc_threshold_cond} can be neglected. In contrast, if the radiation pressure impact is much larger than any heating induced mechanical forces, the term proportional to $\theta$ can be neglected. By demanding that the threshold optical power is minimal, we find that
\begin{subequations}
\begin{align}
	x_\text{0th} & \approx \left\{
	\begin{array}{ll}
		-\frac{\Gamma}{2\sqrt{3}} & :\theta\gg \frac{\beta}{\omega_0}\nu I_\text{max}\\
		\\
		-\frac{\Gamma}{2\sqrt{5}} & :\theta\ll \frac{\beta}{\omega_0}\nu I_\text{max}
	\end{array}
	\right.,
\end{align}
and
\begin{equation}
	I_\text{th} \approx \left\{
	\begin{array}{ll}
		\frac{\omega_0}{Q}\frac{\kappa^2+\omega_0^2}{\eta\theta}\frac{4\Gamma}{3\sqrt{3}} & :\theta\gg \frac{\beta}{\omega_0}\nu I_\text{max}\\
		\\
		\sqrt{\frac{\omega_0^2}{Q}\frac{\kappa^2+\omega_0^2}{\eta\beta\nu} \frac{27}{25\sqrt{5}}\Gamma} & :\theta\ll \frac{\beta}{\omega_0}\nu I_\text{max}
	\end{array}
	\right..
\end{equation}
\end{subequations}
This threshold is shown in Fig.~\ref{fig:small_osc_gamma}, and in Fig.~\ref{fig:small_osc_dissipation_domains} together with different stability regions.

In order to better illustrate the changes in the linear damping coefficient $\gamma$ due to coupling to an optical resonance cavity, we choose a set of realistic parameters, which are given in Table~\ref{table:numerical_parameters}, and draw the resulting $\gamma$ coefficient for a range of $x_0$ and $I_\text{max}$ values. The result is presented in Fig.~\ref{fig:small_osc_gamma}.

\begin{table}[ht]
	\centering
	\renewcommand{\arraystretch}{2}
	\begin{tabular}{ c | c | c  }
		\hline
		parameter & value & units \\[0.5ex]
		\hline \hline
		m & $20\times 10^{-12}$ & \kg \\[0.5ex]
		${\displaystyle \frac{\omega_0}{2\pi}}$ & $160$ & \kilohertz \\[0.5ex]
		Q & $2.5\times 10^5$ &  \\[0.5ex]
		$\alpha_3$ & $3\times 10^{24}$ & ${\displaystyle\frac{1}{\meter^2\sec^2}}$ \\[0.5ex]
		$\gamma_3$ & $9\times 10^{16}$ & ${\displaystyle\frac{1}{\meter^2\sec}}$ \\[0.5ex]
		$\kappa$ & $7.3\times 10^3$ & ${\displaystyle\frac{1}{\sec}}$ \\[0.5ex]
		$\beta$ & $0.001\omega_0 \approx  10^3$ & ${\displaystyle\frac{\radian}{\sec\kelvin}}$ \\[0.5ex]
		$\eta$ & $7.5\times 10^6$ & ${\displaystyle\frac{\kelvin}{\sec\watt}}$ \\[0.5ex]
		$\nu$ & $325$ & ${\displaystyle\frac{\sec}{\kg\:\meter}}$ \\[0.5ex]
		$\theta$ & $4.7$ & ${\displaystyle\frac{\newton}{\kg\:\kelvin}}$ \\[0.5ex]
		$L$ & 0.775 & \micm \\[0.5ex]
		$\Gamma$ & $0.12L=0.093$ & \micm \\[0.5ex]
		$T_0$ & 77 & \kelvin \\[0.5ex]
		$k_\text{max}$ & 25 & \\[0.5ex]
		\hline
	\end{tabular}
	\caption{Values of parameters in a numerical example used to illustrate the results of Sections~\ref{sec:large_osc_theory} and~\ref{sec:small_osc_theory}. All the values are of the same order of magnitude as those found in our experiments, which are reported elsewhere \cite{Zaitsev_et_al_11, Zaitsev_et_al_11b}.}
	\label{table:numerical_parameters}
\end{table}

The coupling of the micromechanical oscillator to an optical resonance cavity does not only introduce linear contributions to the equation of motion, but has an impact on the nonlinear behavior of the system as well. The evolution equation~\eqref{eq:small_osc_complex_evolution_eq} is characteristic for a Duffing-type oscillator with nonlinear damping \cite{Nayfeh_Mook_book_95, Lifshitz&Cross_08, Zaitsev_et_al_11}. The nonlinear coefficients in Eq.~\eqref{eq:small_osc_complex_evolution_eq}, i.e., $q$ and $r$, are functions of the second derivative of the optical power $I(x)$ with respect to displacement.

It follows from Eqs.~\eqref{eq:small_osc_q_and_r} that if $I(x)$ is convex near $x=0$, namely $I_0''>0$, then the nonlinear elastic parameter $q$ is reduced (softening behavior), and the nonlinear dissipation is increased [see Eq.~\eqref{eq:small_osc_r}] if compared to the purely mechanical value $r=\gamma_3/2$. In contrast, if $I(x)$ is concave in the vicinity of $x=0$, namely $I_0''<0$, then the nonlinear elastic parameter is increased (hardening behavior), and the nonlinear dissipation is reduced. At optical powers high enough, the nonlinear dissipation can become negative, suggesting the existence of a large amplitude limit cycle in the system (see Fig.~\ref{fig:small_osc_dissipation_domains}).

Using Eqs.~\eqref{eq:I_approx_parameters}, it can be can be shown that the effective nonlinear corrections to the mechanical equation of motion discussed above change sign when
\begin{subequations}
\begin{align}
	I_0'' & =0,\\
	I_0' & \approx \pm \frac{3\sqrt{3}I_\text{max}}{4\Gamma},\\
	x_0 & \approx \pm\frac{\Gamma}{2\sqrt{3}}.
\end{align}
	\label{eq:I_inflection_point}
\end{subequations}
In this case, $x=0$ is one of the inflection points of $I(x)$.

The magnitude of nonlinear effects in this system strongly depends on the ratio between the thermal relaxation rate $\kappa$ and the mechanical resonance frequency $\omega_0$. At very fast thermal relaxation rates, the elastic coefficient is $q \to 3\alpha_3/2\omega_0$, and also $r\to \gamma_3/2$. As expected, the heating dependent nonlinear terms become negligible when $\kappa\gg\omega_0$, which is a special case of the general result discussed in Sec.~\ref{sec:large_kappa}.

At low values of $\kappa$, i.e., when the thermal relaxation time is significantly smaller than the mechanical oscillation period $2\pi/\Omega_s$, care should be taken when applying the results of the previous section, because the requirement that $\Delta\omega_{0s}\ll\omega_0$ can be easily violated, making the Eq.~\eqref{eq:small_osc_polar_evolution_eq} and all the results following it in Sec.~\ref{sec:small_osc_theory} inapplicable.

%------------------------------------------------------------------------------------------------------------------------------------
	\subsection{Transient behavior}
	\label{sec:small_osc_transient_self_osc}

In order to demonstrate the complex dissipative behavior of our system, we consider the non excited ($f_m=0$, $\sigma_0=0$) solution of Eqs.~\eqref{eq:small_osc_polar_evolution_eq}, which can be written as
\begin{subequations}
\begin{align}
	& \dot A_{1s} +\gamma A_{1s}=-\frac14 rA_{1s}^3, \label{eq:small_osc_limit_cycle_dotA1s}\\
	& \dot\phi+\left(\Delta\omega_s-\frac{q}{4}A_{1s}^2\right)=0. \label{eq:small_osc_limit_cycle_dot_phis}
\end{align}
\end{subequations}
Equation~\eqref{eq:small_osc_limit_cycle_dotA1s} is a regular Bernoulli differential equation, which can be brought to a linear form by a standard transformation $y=A_{1s}^{-2}$. The solution is
\begin{equation}
	A_{1s}^2(t)=\frac{A_{1s}(0)^2e^{-2\gamma t}}{1+\frac{r}{4\gamma_1}A_{1s}(0)^2\left(1-e^{-2\gamma t}\right)},
	\label{eq:small_osc_A1s_ringdown}
\end{equation}
where the initial condition is $A_{1s}(t=0)=A_{1s}(0)$. Equation~\eqref{eq:small_osc_limit_cycle_dot_phis} defines a small correction to the free oscillation frequency.

\begin{figure}
	\centering
	 \includegraphics[width=3.4in]{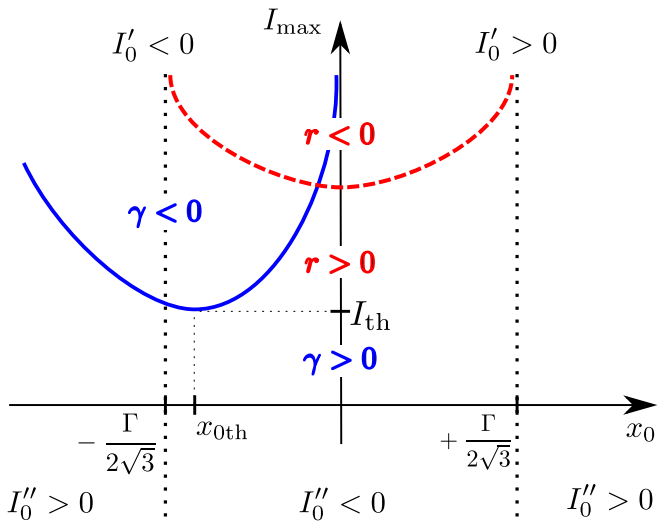}
        \caption{(Color online) Linear and nonlinear dissipation coefficients in an optomechanical resonator and self oscillation thresholds. The black dotted vertical lines limit the area in which $I_0''<0$ [see Eqs.~\eqref{eq:I_inflection_point}]. The solid blue line denotes a self oscillation threshold above which the effective linear damping is negative, i.e., $\gamma<0$. The dashed red line denotes the region in which the nonlinear dissipation is negative ($r<0$), and, therefore, the small amplitude limit cycle $A_\text{LC}$ given in Eq.~\eqref{eq:small_osc_limit_cycle_as} is unstable, suggesting that the existence of an additional large amplitude stable limit cycle is possible.}
        \label{fig:small_osc_dissipation_domains}
\end{figure}

Several interesting cases can be distinguished in Eq.~\eqref{eq:small_osc_A1s_ringdown}.  Figure~\ref{fig:small_osc_dissipation_domains} summarizes all possible cases of linear and nonlinear dissipation as function of the initial displacement $x_0$ and maximal optical power in the cavity $I_\text{max}$.

If the nonlinear dissipation coefficient $r$ is positive, only finite stable solutions of Eq.~\eqref{eq:small_osc_A1s_ringdown} exist. If the linear dissipation coefficient $\gamma$ is also positive, then the system decays almost exponentially to a single steady fixed point $A_{1s}=0$. The rate of decay at times $t>\gamma^{-1}$ is approximately equal to the linear rate $2\gamma$. This decay rate of the optomechanical oscillations can be either larger or smaller than the pure mechanical dissipation rate, $\omega_0/Q+\gamma_3A_{0s}^2$, depending on the sign of $I_0'$ (see also Fig.~\ref{fig:small_osc_gamma}).

In contrast, if $r>0$ but $\gamma<0$ then the system decays not to a trivial zero solution but to a stable limit cycle, whose radius  in the plane of the complex slow changing amplitude $a_s$ \cite{Strogatz_book_94} is given by
\begin{equation}
	|a|^2_\text{LC}=\frac14A^2_\text{LC}=-\frac{\gamma}{r}.
	\label{eq:small_osc_limit_cycle_as}
\end{equation}
The convergence to the limit cycle is again exponential. The result in Eq.~\eqref{eq:small_osc_limit_cycle_as} is correct only if $A_\text{LC}$ is sufficiently small, i.e., if the assumption $A_\text{LC}\ll \Gamma$ holds. The oscillation frequency of this limit cycle can be found from Eq.~\eqref{eq:small_osc_limit_cycle_dot_phis}, resulting in the following expression for the phase variable $\psi$:
\begin{subequations}
\begin{equation}
	\psi_\text{LC}\approx \left(\omega_0-\Delta\omega_s+\frac{q}{4}A_\text{LC}^2\right)t.
\end{equation}
% where
% \begin{equation}
% 	\Delta\omega_\text{tot}=\Delta\omega_s-\frac{q}{4}A_\text{LC}^2.
% \end{equation}
	\label{eq:small_osc_psi_lc}
\end{subequations}
The limit cycle frequency $\omega_0-\Delta\omega_s$ is similar to the one extracted from the local stability analysis of the full dynamical system given in App.~\ref{app:equilibrium_eqs_motion} in the limit $A_\text{LC} \to 0$ [see Eq.~\eqref{eq:app_equilibrium_wH}].

Unlike the unconditionally stable cases described above, the result given in Eq.~\eqref{eq:small_osc_A1s_ringdown} can diverge in finite time if the nonlinear dissipation is negative, i.e., $r<0$. The divergence occurs if the denominator in Eq.~\eqref{eq:small_osc_A1s_ringdown} becomes zero. Here, two cases should be distinguished. If the linear dissipation is positive, i.e., $\gamma>0$, then the system will diverge only if the starting point $A_{1s}(0)>A_\text{LC}$. In other words, the limit cycle described in Eq.~\eqref{eq:small_osc_limit_cycle_as} exists, but is unstable. If, however, both linear and nonlinear dissipation terms are negative - the solution of Eqs.~\eqref{eq:small_osc_limit_cycle_dotA1s} unconditionally diverges. The general large amplitude analysis which is applicable in the last two cases has been presented in Sec.~\ref{sec:large_osc_theory}.

\begin{figure}[!ht]
	\centering
	 \includegraphics[width=3.4in]{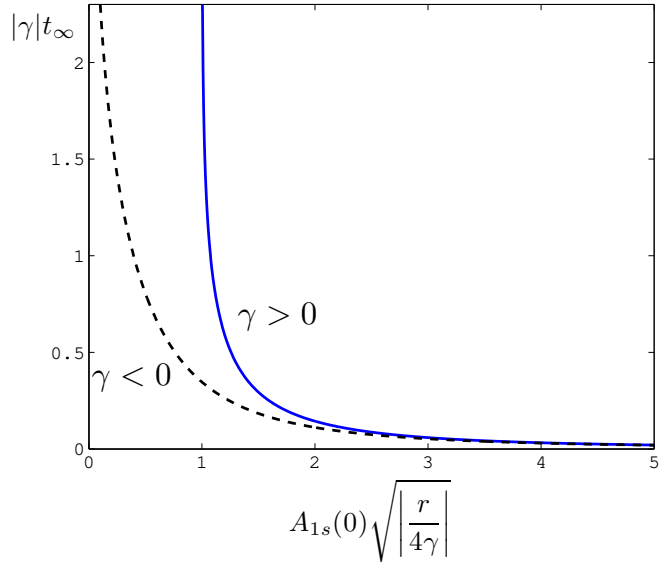}
        \caption{(Color online) Approximate divergence time $t_\infty$ as a function of the initial amplitude $A_{1s}(0)$ [see Eq.~\eqref{eq:small_osc_t_infty}]. The variables are chosen so that the axes are dimensionless ($|\gamma|t_\infty$ vs. $\sqrt{A_{1s}(0)^2|r/4\gamma|}$). Two cases are shown: solid blue line represents the case of positive linear dissipation ($\gamma>0$), dashed black line represents the case of negative linear dissipation ($\gamma<0$). The nonlinear dissipation is negative in both cases ($r<0$).}
        \label{fig:small_osc_t_infty}
\end{figure}

At this point, it is possible to give an estimate of the divergence time $t_\infty$ by requiring that the denominator on the right hand side of Eq.~\eqref{eq:small_osc_A1s_ringdown} vanishes, i.e.,
\begin{equation*}
	1+\frac{r}{4\gamma}A_{1s}(0)^2\left(1-e^{-2\gamma t}\right)=0,
\end{equation*}
resulting in
\begin{equation}
	t_\infty=-\frac{1}{2\gamma}\ln\left(1+\frac{1}{A_{1s}(0)^2}\frac{4\gamma}{r} \right).
	\label{eq:small_osc_t_infty}
\end{equation}
The approximate divergence times according to Eq.~\eqref{eq:small_osc_t_infty} are shown in Fig.~\ref{fig:small_osc_t_infty}. Note that when the absolute value of $\gamma$ is very low, the divergence time $t_\infty$ can be very long if the starting point $A_{1s}(0)$ is close to the unstable limit cycle (for $\gamma>0$) or the origin (for $\gamma<0$). This behavior can be especially important if the system dynamics is simulated numerically, in which case extremely long transient times are undesirable.

% It is important to emphasize that, according to the Poincar\'e-Bendixson theorem \cite{Strogatz_book_94}, no chaotic behavior can occur in this system without external excitation as long as the slow envelope approximation can be employed to reduce the system's behavior to a single complex evolution equation, such as Eq.~\eqref{eq:small_osc_complex_evolution_eq}, limiting the phase space to two dimensional plane. \emph{Review this passage, is it correct at all ???} .

%%%%%%%%%%%%%%%%%%%%%%%%%%%%%%%%%%%%%%%%%%%%%%%%%%%%%%%%%%%%
\section{Self-excited oscillations}
\label{sec:self_osc}

It follows from the stability analysis in the previous Section and in App.~\ref{app:equilibrium_eqs_motion} that a system governed by Eqs.~\eqref{eq:large_osc_polar_evolution_eq} spontaneously develops self-excited oscillations if $\gamma<0$, and can also start self-oscillating if driven far enough from the stable region near the origin in case $\gamma>0$ and $r<0$. Here, we derive the steady state solutions of Eqs.~\eqref{eq:large_osc_polar_evolution_eq} in order to give semi-analytical estimations of the amplitudes of the steady limit cycles that exist in the system and their frequencies.

For convenience, we rewrite Eqs.~\eqref{eq:large_osc_polar_evolution_dotA1},~\eqref{eq:large_osc_Omega} and \eqref{eq:large_osc_A0} for a steady state solution (i.e., $\dot A_1=0$) without external excitation terms below:
\begin{subequations}
\begin{align}
	\Omega & =\omega_0-\frac{\beta\eta}{\kappa}P_0=\omega_0-\Delta\omega_0, \label{eq:self_osc_Omega}\\
	A_0 & = \frac{1}{\Omega^2+\frac32\alpha_3A_1^2}\left[2P_1\beta\eta \frac{\omega_0\kappa}{\kappa^2+\omega_0^2}A_1+P_0 \left(\nu+\frac{\theta\eta}{\kappa}\right)\right],
\end{align}
and
\begin{multline}
	 -\left(\frac{\omega_0}{2Q}+\frac{\gamma_3}{2} A_0^2+2P_2\beta\eta\frac{\omega_0}{\kappa^2+4\omega_0^2}\right)A_1\\
-\frac{\gamma_3}{8}A_1^3-P_1\eta\frac{\omega_0}{\kappa^2+\omega_0^2}\left(2\beta A_0+\frac{\theta}{\omega_0}\right)=0.
	\label{eq:self_osc_dotA1}
\end{multline}
\end{subequations}
The small frequency correction at a given steady state amplitude $A_1$ can be found from Eq.~\eqref{eq:large_osc_polar_evolution_phi}, resulting in
\begin{multline}
	\phi=\Bigg\{-\Delta\omega_0+\frac{3\alpha_3}{2\omega_0}\left(A_0^2+\frac14 A_1^2\right)-P_2\beta\eta \frac{\kappa}{\kappa^2+4\omega_0^2}\\
-\frac{P_1}{A_1}\left[\eta\frac{\kappa}{\kappa^2+\omega_0^2}\left(2\beta A_0+\frac{\theta}{\omega_0}\right)+\frac{\nu}{\omega_0}\right]\Bigg\}t\\
	=-\Delta\omega t,
	\label{eq:self_osc_phi}
\end{multline}
which corresponds to $\Delta\omega_s$ at small amplitudes [see Eq.~\eqref{eq:small_osc_omegas}].

\begin{figure}
	\centering
	 \includegraphics[width=3.4in]{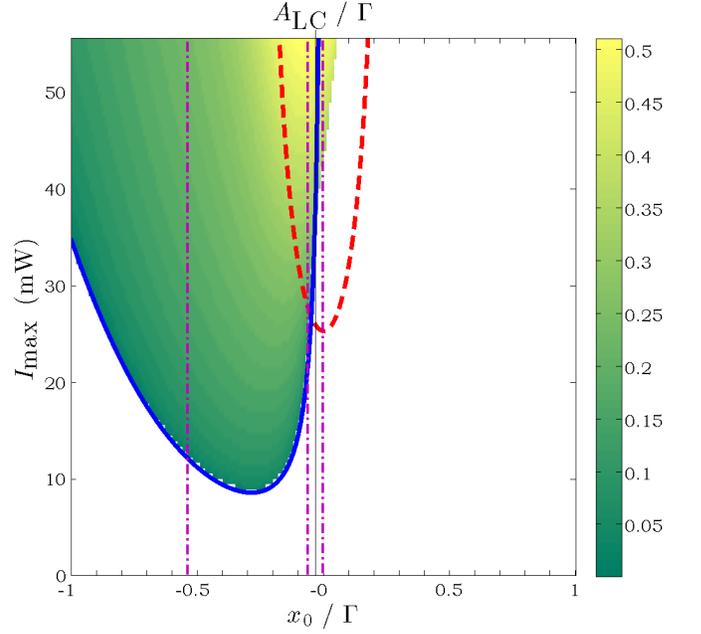}
        \caption{(Color online) The stable limit cycle amplitude vs. the spatial cavity detuning $x_0$ and the optical power $I_\text{max}$. Plotted is the non zero solution of Eq.~\eqref{eq:self_osc_dotA1}, normalized by the width of the optical power peak, $\Gamma$. The parameters of the optomechanical system are given in Table~\ref{table:numerical_parameters}. Similarly to Fig.~\ref{fig:small_osc_dissipation_domains}, the linear dissipation is negative ($\gamma<0$) above the solid blue line, and the nonlinear dissipation is negative ($r<0$) above the dashed red line. The thin dash-dotted magenta lines represent the three values of $x_0$ at which the steady state amplitudes vs $I_\text{max}$ are plotted in Figs.~\ref{fig:theory_self_osc_x0=-0.5Gamma}, \ref{fig:theory_self_osc_x0=-0.02Gamma}, and \ref{fig:theory_self_osc_x0=+0.02Gamma}. The corresponding spatial cavity detuning values are $x_0/\Gamma=-0.5,\;-0.02,\;\text{and } +0.02$.}
        \label{fig:theory_self_osc_vs_x0_Imax}
\end{figure}

In order to illustrate the various possible limit cycles that can occur in a system whose parameters are given in Table~\ref{table:numerical_parameters}, we plot the non zero solutions of Eq.~\eqref{eq:self_osc_dotA1} for a representative range of the mechanical cavity detuning $x_0$ and the optical power $I_\text{max}$ in Fig.~\ref{fig:theory_self_osc_vs_x0_Imax}.

As can be seen in Fig.~\ref{fig:theory_self_osc_vs_x0_Imax}, a limit cycle with non zero amplitude always exists when $\gamma<0$, but only exists for the higher values of optical power when $r<0$. This can be explained by the fact that when the nonlinear dissipation coefficient $r$ is already negative but close to zero, the limit cycle amplitude given by Eq.~\eqref{eq:small_osc_limit_cycle_as} is extremely large, and the small amplitude analysis is inapplicable, as explained in Sec.~\ref{sec:small_osc_transient_self_osc}. In other words, Eq.~\eqref{eq:self_osc_dotA1} can have only the trivial zero solution even when the nonlinear dissipation is negative, but still close to zero.

\begin{figure}
	\centering
	 \includegraphics[width=3.4in]{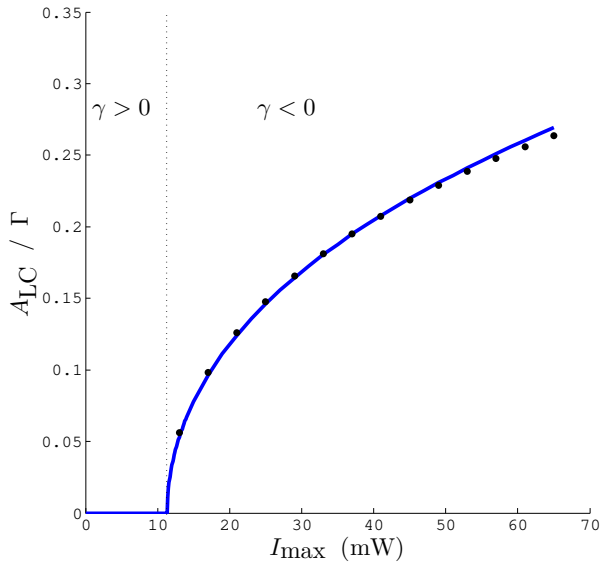}
        \caption{(Color online) Steady state amplitude as function of the optical power $I_\text{max}$ at $x_0=-0.5\Gamma$. The optomechanical system's parameters are given in Table~\ref{table:numerical_parameters}. The plot corresponds to a cross section of Fig.~\ref{fig:theory_self_osc_vs_x0_Imax}, which is defined there by the leftmost dash-dotted magenta line. The large black dots are the estimations of the stable limit cycle for small amplitudes as given by Eq.~\eqref{eq:small_osc_limit_cycle_as}.}
        \label{fig:theory_self_osc_x0=-0.5Gamma}
\end{figure}

The steady state solution of Eq.~\eqref{eq:self_osc_dotA1} for $x_0=-0.5\Gamma$ is shown in Fig.~\ref{fig:theory_self_osc_x0=-0.5Gamma}. The zero solution is stable as long as the linear dissipation is positive, and a small stable limit cycle develops when $\gamma$ becomes negative.

\begin{figure}
	\centering
	 \includegraphics[width=3.4in]{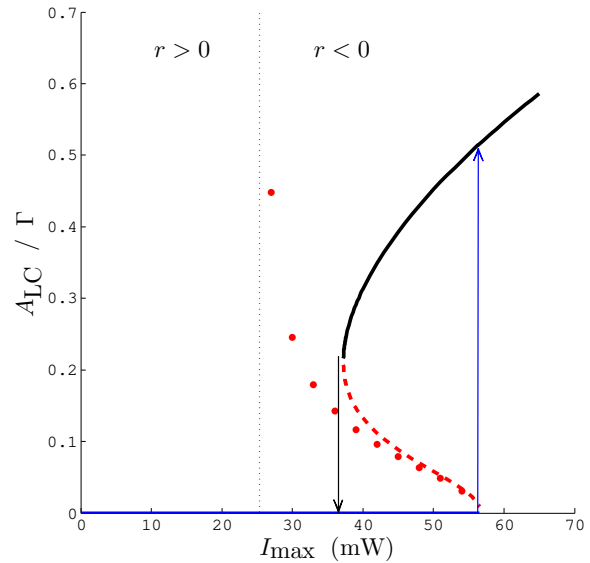}
        \caption{(Color online) Steady state amplitudes as functions of the optical power $I_\text{max}$ at $x_0=-0.02\Gamma$. The optomechanical system's parameters are given in Table~\ref{table:numerical_parameters}. The plot corresponds to a cross section of Fig.~\ref{fig:theory_self_osc_vs_x0_Imax}, which is defined there by the middle dash-dotted magenta line. The solid lines correspond to a stable steady limit cycle (black) and a stable zero solution (blue). The dashed red line corresponds to the unstable fixed point (i.e., separatrix).  The system is bistable in a certain range of optical powers, whose limits are marked by thin vertical arrows. The large red dots are the estimations of the unstable limit cycle for small amplitudes as given by Eq.~\eqref{eq:small_osc_limit_cycle_as}.  As expected, the red dots align well with the unstable solution of Eq.~\eqref{eq:self_osc_dotA1} at lower values of $A_\text{LC}$.}
        \label{fig:theory_self_osc_x0=-0.02Gamma}
\end{figure}

It is interesting to compare the case above, in which the nonlinear dissipation is positive when the zero solution loses stability (see Fig.~\ref{fig:theory_self_osc_x0=-0.5Gamma}), with a case in which, as the optical power increases, the nonlinear dissipation becomes negative before the linear dissipation does. Such a case for $x_0=-0.02\Gamma$ is presented in Fig.~\ref{fig:theory_self_osc_x0=-0.02Gamma}. As can be seen in this figure, two stable solutions and one unstable solution coexist in a bistable region, whose limits are marked by vertical arrows. This results in an amplitude hysteresis when the optical power $I_\text{max}$ or the spatial detuning $x_0$ are swept.

The linear damping in the bistable region is positive, and, therefore, the zero solution remains stable. In addition to the zero solution, another large amplitude stable solution exists, because the nonlinear damping coefficient $r$ is negative. At small amplitudes, the amplitude of the separatrix, denoted by the dashed red line, corresponds to the solution of Eq.~\eqref{eq:small_osc_limit_cycle_as}, which is marked by large red dots. At optical powers high enough, the linear damping becomes negative, the separatrix amplitude reaches zero, and the only remaining stable solution is the large amplitude limit cycle.

\begin{figure}
	\centering
	 \includegraphics[width=3.4in]{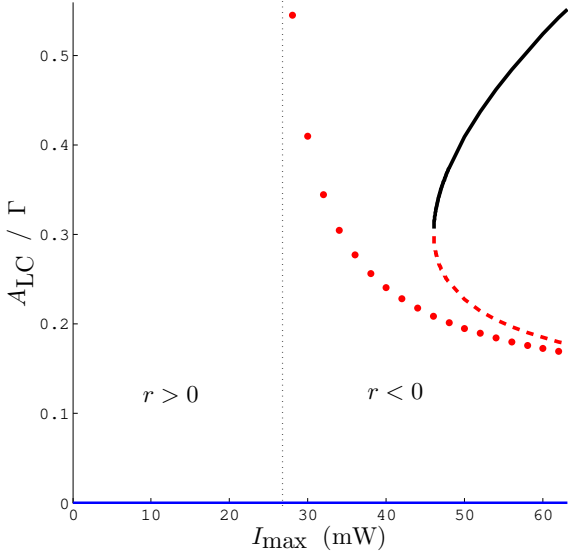}
        \caption{(Color online) Steady state amplitudes as functions of the optical power $I_\text{max}$ at $x_0=+0.02\Gamma$. The optomechanical system's parameters are given in Table~\ref{table:numerical_parameters}. The plot corresponds to a cross section of Fig.~\ref{fig:theory_self_osc_vs_x0_Imax}, which is defined there by the rightmost dash-dotted magenta line. The solid lines correspond to a stable limit cycle (black) and a stable zero solution (blue). The dashed red line corresponds to the unstable limit cycle (i.e., separatrix).  The system is bistable above a certain optical power. The large red dots are the estimations of the unstable limit cycle for small amplitudes as given by Eq.~\eqref{eq:small_osc_limit_cycle_as}.  As expected, the red dots align well with the unstable solution of Eq.~\eqref{eq:self_osc_dotA1} at lower values of $A_\text{LC}$.}
        \label{fig:theory_self_osc_x0=+0.02Gamma}
\end{figure}

The third typical configuration of limit cycles in this system is presented in Fig.~\ref{fig:theory_self_osc_x0=+0.02Gamma}, where a case for $x_0=+0.02\Gamma$ is shown. Here, the linear damping is unconditionally positive, therefore, the zero solution is always stable. In addition, when the nonlinear damping is negative, another couple of limit cycles can exist with finite amplitudes, an unstable one, acting as a separatrix, and a stable one.

\begin{figure}
	\centering
	 \includegraphics[width=3.4in]{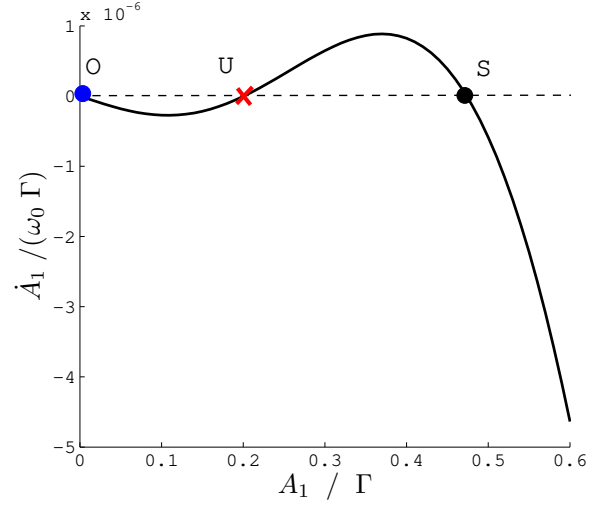}
        \caption{(Color online) The slow envelope velocity, $\dot A_1$ [see Eq.~\eqref{eq:large_osc_polar_evolution_dotA1}], as a function of the amplitude $A_1$ at the bistable region shown in Fig.~\ref{fig:theory_self_osc_x0=+0.02Gamma}. The spatial optical cavity detuning is $x_0=+0.02\Gamma$, the optical power is $I_\text{max}=55\milliwatt$, and the optomechanical system's parameters are given in Table~\ref{table:numerical_parameters}. In order to plot a dimensionless function, the velocity $\dot A_1$ is normalized by the characteristic fast mechanical velocity $\omega_0\Gamma$. The stable zero amplitude solution $O$ is denoted by a blue dot. The unstable limit cycle $U$ is denoted by a red cross, and the stable finite amplitude limit cycle $S$ is denoted by a black dot.}
        \label{fig:theory_dotA1_example}
\end{figure}

In order to complete the picture of the different limit cycles which are possible in the optomechanical system under study, the slow envelope velocity, $\dot A_1$ [see Eq.~\eqref{eq:large_osc_polar_evolution_dotA1}], is drawn  in Fig.~\ref{fig:theory_dotA1_example} as a function of the amplitude $A_1$ at the bistable region shown in Fig.~\ref{fig:theory_self_osc_x0=+0.02Gamma}.

Several features of Fig.~\ref{fig:theory_dotA1_example} and Eq.~\eqref{eq:large_osc_polar_evolution_dotA1} should be emphasized. First, the stable finite amplitude solution $S$ is separated from the stable zero solution $O$ by the unstable solution $U$. Second, the pair of fixed points $U$ and $S$ appear in a saddle node bifurcation when the optical power is increased (in the case shown in Fig.~\ref{fig:theory_dotA1_example}, this bifurcation has already happened). Third, the positive mechanical nonlinear damping, i.e., $\gamma_3>0$, is prevalent at large amplitudes, driving the slow envelope velocity $\dot A_1$ to large negative values, and, therefore, preventing the existence of any other limit cycles with larger amplitudes. If the nonlinear mechanical effects are negligible, the system can become multistable, with several coexisting large amplitude limit cycles \cite{Hane&Suzuki_96, Metzger_et_al_08}.

% 	\subsection{Mechanical noise in the slow envelope approximation}
% 	\label{sec:noise_slow_envelope}

% \emph{Mechanical noise power is a function of temperature and, therefore, a function of displacement. Calculate the averaged noise power in evolution equation, discuss apparent "mode cooling" in case $Q_\text{rp}>0$. Maybe, discuss thermal escape from steady zero state to the small limit cycle state for $qp>0$, $\gamma_1<0$. Another possible subject - significant noise squeezing in the Duffing-like equation of motion due to "mode cooling".}
% 	

%%%%%%%%%%%%%%%%%%%%%%%%%%%%%%%%%%%%%%%%%%%%%%%%%
\section{Numerical validation and the limits of accuracy}
\label{sec:numerical_validation}

In order to validate the analytical expressions derived above in Eqs.~\eqref{eq:large_osc_polar_evolution_eq}, \eqref{eq:large_osc_Omega}, and \eqref{eq:large_osc_A0}, we compare them to the results of the direct numerical integration of Eqs.~\eqref{eq:mech_osc_orig}, \eqref{eq:omega_m}, and \eqref{eq:temp_diff_eq}. The values of all parameters used in the numerical simulation are given in Table~\ref{table:numerical_parameters}. The value of the optical power $I(x)$ in numerical simulations is calculated exactly, i.e., $k_\text{max}=\infty$. The numerical integrations were done using the Matlab software.

\begin{figure}
	\centering
	 \includegraphics[width=3.4in]{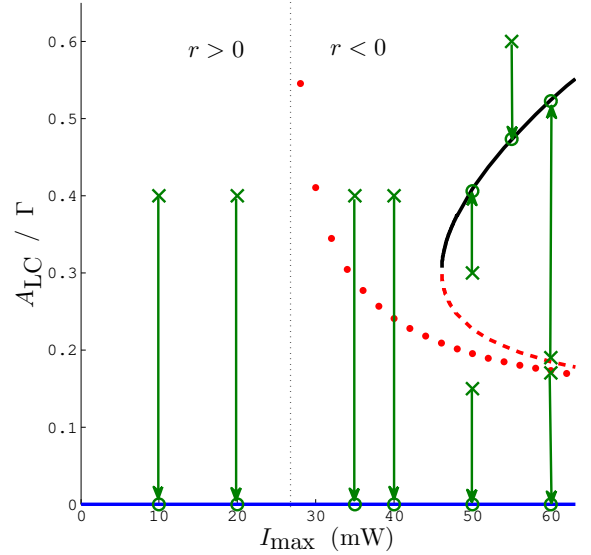}
        \caption{(Color online) Numerical validation of the slow envelope approximation results in Sec.~\ref{sec:large_osc_theory}. The limit-cycle amplitudes as given by the solutions of Eq.~\eqref{eq:self_osc_dotA1} are compared to the results of a full numerical integration of Eq.~\eqref{eq:mech_osc_orig}. The optomechanical system's parameters and the notation are similar to those used in Fig.~\ref{fig:theory_self_osc_x0=+0.02Gamma}. In addition, the initial conditions for the numerical simulations (green crosses) are shown, connected by thin green arrows to the final numerical solutions (green circles).}
        \label{fig:theory_num_validation_ALC}
\end{figure}

The numerical results for the stable limit cycle amplitudes at $x_0=+0.02\Gamma$ are shown in Fig.~\ref{fig:theory_num_validation_ALC}, together with the semi-analytical (i.e., slow envelope approximation) results already presented in Fig.~\ref{fig:theory_self_osc_x0=+0.02Gamma}. The comparison yields good agreement.

\begin{figure}
	\centering
	 \includegraphics[width=3.4in]{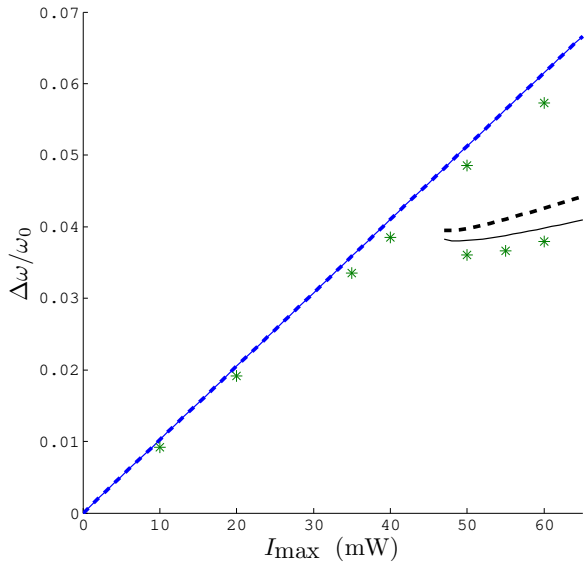}
        \caption{(Color online) Numerical validation of the slow envelope approximation results for frequency shifts from the mechanical frequency $\omega_0$. Note that the frequency shift is defined as $\omega_0-\Delta\omega$, i.e., the oscillation frequency is reduced when $\Delta\omega$ is positive. The semi-analytical limit-cycle and free small oscillation frequency shifts as given by the solutions of Eqs.~\eqref{eq:self_osc_Omega} and~\eqref{eq:self_osc_phi} are compared to the frequency shifts extracted from the results of a full numerical integration of Eq.~\eqref{eq:mech_osc_orig}. The optomechanical system's parameters are similar to those used in Fig.~\ref{fig:theory_self_osc_x0=+0.02Gamma}  and are given in Table~\ref{table:numerical_parameters}. Dashed lines represent the frequency shift which is solely due to the averaged heating (Eq.~\eqref{eq:self_osc_Omega}) for a large limit cycle (black segment on the right) and small vibrations near the origin (blue almost diagonal line across the figure). Thin solid lines of the same colors represent the more exact solution, which incorporates both Eqs.~\eqref{eq:self_osc_Omega} and~\eqref{eq:self_osc_phi}. The results of numerical simulations are represented by green asterisks.}
        \label{fig:theory_num_validation_Omega}
\end{figure}

The slow envelope approximation gives an estimation of the oscillation frequencies associated with large limit cycles [see Eqs.~\eqref{eq:self_osc_Omega} and~\eqref{eq:self_osc_phi}] and their small vibration limit [see Eqs.~\eqref{eq:small_osc_psi_lc}]. In Fig.~\ref{fig:theory_num_validation_Omega}, the free oscillation frequencies extracted from the numerical integration results are compared with the semi-analytical results given in Eqs.~\eqref{eq:self_osc_Omega} and~\eqref{eq:self_osc_phi} for $x_0=+0.02\Gamma$.

The limit cycle oscillation frequencies calculated using Eqs.~\eqref{eq:self_osc_Omega} and~\eqref{eq:self_osc_phi} have a reasonable accuracy only when $\Delta\omega \lesssim 0.1\omega_0$. This is due to the fact that we have neglected terms proportional to powers higher than one of $\Delta\omega_0$ in Eqs.~\eqref{eq:large_osc_polar_evolution_eq} and \eqref{eq:large_osc_A0}. This assumption of small frequency shift becomes increasingly inaccurate at high optical powers, as can be seen in Fig.~\ref{fig:theory_num_validation_Omega}.

In general, the linear expression in Eq.~\eqref{eq:omega_m} is valid for small frequency corrections and for small temperature changes only. The accurate relation between the mechanical frequency and the effective temperature is usually more complicated, and strongly depends on the specific mirror configuration. For example, if a uniform doubly clamped beam with high internal tension is used as a mirror, its fundamental mode frequency can be approximated by a frequency of a fundamental harmonic of a pure string \cite{Pandey_et_al_10}
\begin{equation*}
	\omega_\text{string}(T) =\frac{\pi}{L}\sqrt{\frac{S(T)}{m/L}},
\end{equation*}
where
\begin{equation*}
	S(T) \approx S_0-E\alpha(T-T_0),
\end{equation*}
and $\omega_\text{string}$ is the string's angular vibration frequency, $m$ and $L$ are the mass and the length of the string respectively, $S(T)$ is the temperature dependent total tension in the string, $S_0$ is the tension at $T=T_0$, $E$ is the Young's modulus, and $\alpha$ is the thermal linear expansion coefficient. Here, it is assumed that both the difference between the relaxed beam length and its actual length, and the change in the spring's tension due to heating are small. In addition, the Young's modulus and the thermal expansion coefficient are assumed to be constant in the relevant range of temperatures. One should also remember that the notion of a single effective temperature $T$ may not be sufficient to describe the thermally dependent mechanical behavior of a complex micromechanical structure.

Another limit on the accuracy of the model described in Sec.~\ref{sec:large_osc_theory} stems from the small nonlinearity assumption made in the slow envelope approximation \cite{Nayfeh_book_81}. Specifically, only if the contribution of the nonlinear terms in Eq.~\eqref{eq:mech_osc_orig} is much smaller than the magnitude of the linear terms in the same equation, i.e., only if
\begin{equation*}
	\alpha_3A^2 \ll \omega_0^2,
\end{equation*}
and
\begin{equation*}
	\gamma_3A^2 \ll \frac{\omega_0}{Q},
\end{equation*}
then the harmonic solution assumption in Eq.~\eqref{eq:large_osc_x} together with the averaging process used in Sec.~\ref{sec:large_osc_theory} are valid.

%%%%%%%%%%%%%%%%%%%%%%%%%%%%%%%%%%%%%%%%%%%%%%%%%%%%%%%%%%%%%
\section{Summary}
\label{sec:summary}
A coupling between an optical resonance cavity and a micromechanical resonator presents an interesting challenge for building a simple yet comprehensive model, which is able to capture the complicated dynamics of the coupled system in a small set of relatively simple equations of motion. In this work, we have created such a model for a low finesse optomechanical resonance cavity in which the elastic element is realized in the form of a vibrating nonlinear micromechanical mirror.

The optomechanical cavity is assumed to be constantly pumped by monochromatic laser light. Due to the low finesse of the cavity, the optical response time is considered to be very fast compared to the mechanical resonance frequency, and, therefore, the optical power inside the cavity can be described as an instantaneous function of the mirror's displacement [see Eq.~\eqref{eq:opt_power_periodic_Lorentzian}]. Under these assumptions, we write a set of coupled differential equations which describe the mechanical and thermal dynamics of the system [see Eqs.~\eqref{eq:mech_osc_orig} and~\eqref{eq:temp_diff_eq}, respectively].

The optical power influences the micromechanical mirror's dynamics both directly in the form of radiation pressure, and indirectly through heating. Radiative heating causes the mechanical resonance frequency to change [see Eq.~\eqref{eq:omega_m}]. In addition, a direct thermal force can exist in a mirror in the form, for example, of a bimorph thermal actuation [see Eq.~\eqref{eq:Fth}]. The important property of all heating dependent forces is the retardation that they introduce into the equations of motion, which results in changes in the effective dissipation in the micromechanical system.

The micromechanical mirror itself is described as a Duffing-like weakly nonlinear oscillator with nonlinear (cubic) dissipation. The motion of the mirror can be approximated by a simple harmonic function with slow varying amplitude and phase. Averaging over a single "fast" period of mechanical oscillation results in a set of slow evolution equations for the slow varying amplitude and phase. These equations are given for the externally excited case in Sec.~\ref{sec:large_osc_theory}, and for the case in which no external excitation exists - in Sec.~\ref{sec:self_osc}. In addition, estimations of the oscillation frequency and the static deflection are derived in Sec.~\ref{sec:large_osc_theory}.

Unfortunately, the full evolution equations for arbitrary amplitudes do not have a simple analytical solution. However, they do have a convenient semi-analytical closed form, and can be readily solved by any software designed for numerical calculations, such as the Matlab package used in this work. The solution of the first-order evolution equations requires significantly less computing power than the full numerical integration of the original equations of motion, which can be computationally prohibitive, especially in the case of low damping rates and very long transient times. One must bear in mind, however, that a slow varying envelope approximation of a general dynamic system may have a deficiency of missing additional nonlinear phenomena such as coexisting multi-stable limit cycles, quasi-periodic response (due to incommensurate external and limit-cycle frequencies), homoclinic bifurcations and possible chaos.
%, e.g., homoclinic bifurcations and possible chaos.

The evolution equations can be further simplified if the mechanical amplitude is small. It has been shown in Sec.~\ref{sec:small_osc_theory} that both linear and nonlinear terms originating from the optomechanical coupling can be found in the resulting small amplitude complex evolution equation~\eqref{eq:small_osc_complex_evolution_eq}. The changes in the effective linear and nonlinear dissipation, which are functions both of the spatial cavity detuning and the pumping optical power, are most important [see Eqs.~\eqref{eq:small_osc_gamma_and_Deltaws} and~\eqref{eq:small_osc_q_and_r}]. For example, if the spatial cavity detuning is negative, the effective linear dissipation can become negative at optical powers above a certain threshold, causing a small limit cycle (i.e., self oscillations) to appear. The threshold, the frequency, and the amplitude of these small self oscillations can be predicted with reasonable accuracy using the small amplitude approximation [see Eq.~\eqref{eq:small_osc_limit_cycle_as} and Fig.~\ref{fig:theory_self_osc_x0=-0.5Gamma}]. These results coincide with the predictions of the stability analysis of the full dynamical system which is given in App.~\ref{app:equilibrium_eqs_motion}.

Even when the linear effective damping remains positive, a stable limit cycle with a large amplitude can coexist with a stable zero solution in the region in which the nonlinear damping is negative. In such a case, a hysteresis in the self oscillation amplitude is possible in the system when either the optical power or the spatial cavity detuning are swept back and forth. All the possible situations leading to self oscillations have been summarized in Sec.~\ref{sec:self_osc}.

Finally, we compare the results which are derived from the slow envelope evolution equations with the full numerical integration of the original equations of motion in Sec.~\ref{sec:numerical_validation}. As expected, the semi-analytical results of this work are well-correlated with the full numerical integration results as long as the major assumptions of the slow envelope approximation are satisfied. In other words, the validity of the majority of the results presented here depends on the assumption that all the optical dependent and nonlinear terms in the original equation of motion~\eqref{eq:mech_osc_orig} are small.

In our treatment, the dependence of the different terms in the equation of motion on the effective temperature of the vibrating mechanical element has the simplest, i.e., linear, form. In general, the method of slow envelope and the averaging technique used in this study can be utilized in order to deal with more complex and more realistic relations between the heating and the oscillation frequency or the thermal force. In addition, further development of the ideas presented above may incorporate a treatment of large frequency changes due to heating and a dependence of additional parameters, such as nonlinear elastic coefficient and all mechanical dissipation coefficients, on temperature.

Based on the theory presented here, an experimental study was conducted by us, which was reported elsewhere~\cite{Zaitsev_et_al_11b}. A comparison between the experimental results and the theoretical model developed in this article yields a good agreement. In particular, the quantitative theoretical model successfully predicted the experimentally measured changes in the linear effective damping, the cubic nonlinearities, the threshold of the self oscillations, the frequency and the amplitude of the self oscillations, and the resonance frequency of the micromechanical mirror under different conditions. The experimental study was done using micromechanical mirrors with two different geometries and material compositions.

%%%%%%%%%%%%%%%%%%%%%%%%%%%%%%%%%%%%%%%%%%%%%%%%%%%%%%%
\section{Acknowledgments}

This work is supported by the German Israel Foundation under grant 1-2038.1114.07, the Israel Science Foundation under grant 1380021, the Deborah Foundation, Eliyahu Pen Research Fund, Russell Berrie Nanotechnology Institute, the European STREP QNEMS Project and MAFAT.

%%%%%%%%%%%%%%%%%%%%%%%%%%%%%%%%%%%%%%%%%%%%%%%%%%%%%%%
\appendix

%--------------------------------------------------------------------------------------------------------------------------------
\section{Spatial Fourier series of a periodic optical power function}
\label{app:opt_power_Fourier_series}

In order to calculate an analytical expression for $c_k$ in Sec.~\ref{sec:optical_resonance_cavity}, we proceed as follows. We rewrite Eqs.~\eqref{eq:opt_power_periodic_Lorentzian} and~\eqref{eq:large_osc_I_Fourier}  as
\begin{subequations}
\begin{equation}
	I(x)=\frac{hI_\text{max}}{1+h-\cos y}=\sum_{k=-\infty}^\infty \beta_k e^{jky},
	\label{eq:large_osc_I_for_ck}
\end{equation}
where
\begin{align}
	y & = 2\pi\frac{x-x_0}{L}, \\
	h & = \frac{\pi^2}{2} \left(\frac{\Gamma}{L}\right)^2, \label{eq:ck_h}\\
	\beta_k & =c_ke^{j2\pi k\frac{x_0}{L}}. \label{eq:large_osc_ck_beta_k_vs_ck}
\end{align}
\end{subequations}
Multiplying both sides of Eq.~\eqref{eq:large_osc_I_for_ck} by $1+h-\cos y$, using the fact that $\cos y=(e^{jy}+e^{-jy})/2$, and separating terms corresponding to different harmonics, one finds
\begin{equation}
	(1+h)\beta_k-\frac12(\beta_{k-1}+\beta_{k+1})=\left\{
	\begin{array}{l l}
		hI_\text{max} & : k=0 \\
		0 & : k \neq 0
	\end{array}
	\right. .
	\label{eq:large_osc_beta_k_recursion}
\end{equation}
Note that $\beta_k=\beta_{-k}$, and $\beta_k$ are real because $I(y)$ is a real even function. Assuming that $\beta_k$ can be represented as
\begin{equation}
	\beta_k=I_\text{max}\chi\alpha^{|k|},
	\label{eq:large_osc_beta_k}
\end{equation}
where $\chi$ and $\alpha$ are real, and substituting Eq.~\eqref{eq:large_osc_beta_k} into Eq.~\eqref{eq:large_osc_beta_k_recursion} for positive values of $k$ results in
\begin{equation*}
	\alpha^2-2(1+h)\alpha+1=0.
\end{equation*}
The solution which ensures series convergence by satisfying the condition $0<\alpha<1$ is
\begin{subequations}
\begin{equation}
	\alpha=1+h-\sqrt{(1+h)^2-1}.
	\label{eq:large_osc_ck_alpha}
\end{equation}
The value of $\chi$ can be found from Eq.~\eqref{eq:large_osc_beta_k_recursion} for the case in which $k=0$, giving
\begin{equation}
	\chi=\frac{h}{\sqrt{(1+h)^2-1}}.
\end{equation}
Finally, Eq.~\eqref{eq:large_osc_ck_beta_k_vs_ck} gives
\begin{equation}
	c_k=I_\text{max}\chi\alpha^{|k|}e^{-j2\pi k\frac{x_0}{L}}.
	\label{eq:large_osc_analytic_ck}
\end{equation}
	\label{eq:large_osc_analytic_I_series}
\end{subequations}

% The following paragraph can be substituted by the long discussion below
It is straightforward to show that if the finesse is bigger than unity, i.e., $\mathcal F$ is of order of ten or higher, the truncation error in Eq.~\eqref{eq:large_osc_I_Fourier_kmax} is negligible if $k_\text{max}\gtrsim \mathcal F$.

% An error introduced by truncating the optical power series at some arbitrary $k_\text{max}$ value can be defined as
% \begin{equation}
% 	error=1-\frac{\displaystyle\sum_{k=-k_\text{max}}^{k_\text{max}} |c_k|^2}{\displaystyle\frac{1}{L}\int_0^L |I(x)|^2 dx}.
% 	\label{eq:large_osc_I_series_error}
% \end{equation}
% According to Parseval's theorem,
% \begin{equation*}
% 	error \xrightarrow{k_\text{max}\to\infty} 0.
% \end{equation*}

% It follows from Eqs.~\eqref{eq:ck_h} and~\eqref{eq:finesse} that
% \begin{equation*}
% 	h=\frac{\pi^2}{2\mathcal F^2}.
% \end{equation*}
% Assuming that the finesse is sufficiently large, i.e., $\mathcal F>1$, we can rewrite Eq.~\eqref{eq:large_osc_ck_alpha} as
% \begin{equation*}
% 	\alpha=1-\frac{\pi}{\mathcal F}+O\left(\frac{1}{\mathcal F^2}\right).
% \end{equation*}
% We note that $(1-x)<e^{-x}$ for $0<x<1$. Applying this inequality to the last equation we find that
% \begin{equation*}
% 	\alpha^{|k|} \lesssim e^{-|k|\frac{\pi}{\mathcal F}}.
% \end{equation*}

% In order to give an upper boundary for the error defined in Eq.~\eqref{eq:large_osc_I_series_error}, we use Eq.~\eqref{eq:large_osc_analytic_ck} and write
% \begin{equation*}
% 	error=1-\frac{1+\displaystyle 2\sum_{k=1}^{k_\text{max}} \alpha^{2k}}{1+\displaystyle 2\sum_{k=1}^{\infty} \alpha^{2k}}=\frac{\alpha^{2k_\text{max}+2}}{1+\alpha^2} \lesssim e^{-(2k_\text{max}+2)\frac{\pi}{\mathcal F}}.
% \end{equation*}
% It follows that the truncation error in Eq.~\eqref{eq:large_osc_I_Fourier_kmax} is negligible if $k_\text{max}\gtrsim \mathcal F$.

\begin{figure}
% 	\centering
% 	 \includegraphics[width=3.4in]{Figures/truncated_I_graph.eps}
	 \includegraphics[width=3.4in]{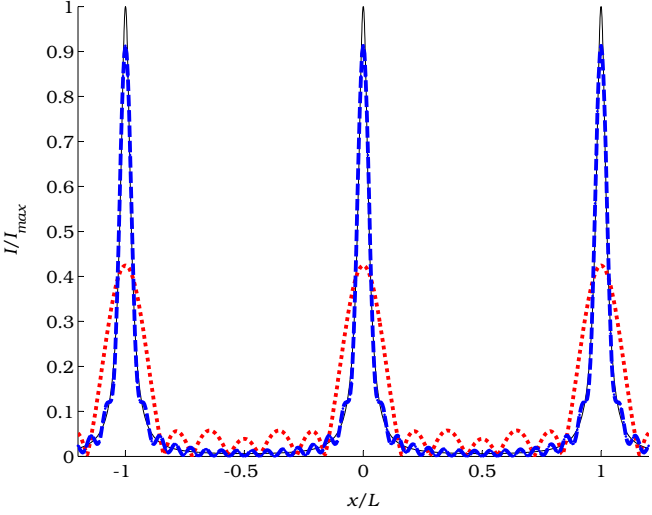}
      \caption{(Color online) Comparison between an exact periodic function describing the optical power in a resonator given in Eq.~\eqref{eq:opt_power_periodic_Lorentzian} (solid black line) and its truncated spatial Fourier series. The optical resonance width is $\Gamma=0.05L$. Red dotted line corresponds to $k_\text{max}=3$. Blue dashed line corresponds to $k_\text{max}=15$.}
      \label{fig:large_osc_I_series}
\end{figure}

An example of several truncated Fourier series calculated using Eqs.~\eqref{eq:large_osc_analytic_I_series} for different values of $k_\text{max}$ is shown in Fig.~\ref{fig:large_osc_I_series}.

%------------------------------------------------------------------------------------------------------------------------------------

\section{Equilibrium analysis of the equations of motion}
\label{app:equilibrium_eqs_motion}

In this section, we analyze the equilibrium position of the third order autonomous nonlinear dynamical system defined by Eqs.~\eqref{eq:mech_osc_orig} and \eqref{eq:temp_diff_eq} where the external exciting force is zero ($f_m=0$).

By defining new variables $p=\dot x$ and $\Delta T=T-T_0$, the equations of motion can be rewritten as
\begin{subequations}
\begin{align}
\dot x & = p,\\
\dot p & = -\left(\frac{\omega_0}{Q}+\gamma_3x^2\right)p -\omega_m^2x-\alpha_3x^3+\theta\Delta T+\nu I(x),\\
\Delta\dot T & = -\kappa\Delta T +\eta I(x),
\end{align}
	\label{eq:app_equilibrium_dyn_sys_3eqs}%
\end{subequations}
where parameters defined in Sec.~\ref{sec:eq_of_motion} have been used.

The equilibrium position of the dynamical system (i.e., the fixed point) is readily obtained by setting the velocities (i.e., the left hand side of Eqs.~\eqref{eq:app_equilibrium_dyn_sys_3eqs}) to zero. This results in a transcendental function for the equilibrium displacement $A_{0s}$,
\begin{subequations}
\begin{equation}
	\Omega_s^2A_{0s}+\alpha_3A_{0s}^3-\nu I(A_{0s})-\theta\Delta T_0=0,
	\label{eq:app_equilibrium_A0s_eq}%
\end{equation}
where the equilibrium temperature shift $\Delta T_0$ is
\begin{equation}
	\Delta T_0=\frac{\eta}{\kappa}I(A_{0s}) ,
	\label{eq:app_equilibrium_DT0}%
\end{equation}
and the equilibrium mechanical resonance frequency is
\begin{equation}
	\Omega_s=\omega_0-\beta\Delta T_0.
	\label{eq:app_equilibrium_Omegas}
\end{equation}
	\label{eq:app_equilibrium_fixed_point_params}%
\end{subequations}
In the limit of a very small equilibrium displacement $A_{0s}\approx 0$ (i.e., the limit of very weak optomechanical forces), the Eqs.~\eqref{eq:app_equilibrium_fixed_point_params} converge to the similar equations~\eqref{eq:small_osc_params} derived in Sec.~\ref{sec:small_osc_theory}.

In general, multiple solutions of Eqs.~\eqref{eq:app_equilibrium_fixed_point_params} may co-exist, corresponding to several stable and unstable fixed points under the same experimental conditions. However, in the case in which the thermal frequency shift, the radiation pressure and the thermal force are all considered small, the limiting case of Eqs.~\eqref{eq:small_osc_params} predicts a single stable fixed point with a small static displacement $A_{0s}\ll\Gamma$.

Stability of the equilibrium is obtained via a local perturbation of the system fixed point defined by Eqs.~\eqref{eq:app_equilibrium_fixed_point_params}, resulting in a linear variation
\begin{equation*}
	\left(\begin{array}{c}
	\dot x \\
	\dot p \\
	\Delta\dot T
	\end{array}\right)
	=
	M
	\left(\begin{array}{c}
	x-A_{0s} \\
	p \\
	\Delta T-\Delta T_0
	\end{array}\right),
\end{equation*}
where $M$ is the Jacobian matrix of the first derivatives of the system functions given by the right hand parts of Eqs.~\eqref{eq:app_equilibrium_dyn_sys_3eqs}. Thus, equilibrium stability can readily be obtained by evaluating the eigenvalues $\lambda_1$, $\lambda_2$ and $\lambda_3$ of $M$, which satisfy:
\begin{equation*}
	\lambda^3+c_1\lambda^2+c_2\lambda+c_3=0,
	\label{eq:app_equilibrium_character_polynom}
\end{equation*}
where
\begin{subequations}
\begin{align}
	c_1 & = \kappa+\frac{\omega_0}{Q}+\gamma_3A_{0s}^2,\\
	c_2 & = \kappa\left(\frac{\omega_0}{Q}+\gamma_3A_{0s}^2\right)+\Omega_s^2+3\alpha_3A_{0s}^2-\nu I'(A_{0s}),\\
	c_3 & = \kappa\left(\Omega_s^2+3\alpha_3A_{0s}^2\right)-\left[\kappa\nu+\eta(2\beta\Omega_sA_{0s}+\theta)\right]I'(A_{0s}),%
\end{align}
	\label{eq:app_equilibrium_ci}%
\end{subequations}
and where a prime denotes differentiation with respect to the mechanical displacement $x$.

Asymptotic stability of the equilibrium (i.e., $\Real \{\lambda_i\}<0$) is defined by positive coefficients and a positive second Hurwitz determinant, namely, $c_i>0$ and $\Delta_2=(c_1c_2-c_3)>0$. Loss of equilibrium stability is defined by a zero eigenvalue ($c_3=0$), or a Hopf bifurcation where the Jacobian matrix $M$ has a pair of pure imaginary eigenvalues, i.e., $\lambda_{1,2}=\pm i\omega_H$.

The zero eigenvalue condition $c_3=0$ can be rewritten in a differential form as
\begin{multline*}
	\left(\Omega_s^2+3\alpha_3A_{0s}^2\right)dx+2A_{0s}\Omega_s\frac{\beta\eta}{\kappa}dI(A_{0s})\\
	=\left(\nu+\frac{\eta\theta}{\kappa}\right)dI(A_{0s}).
\end{multline*}
This equation can be readily understood as a condition of equality between the thermally dependent nonlinear elastic force (left hand side terms) and the optomechanical forces (right hand side terms). This condition describes a saddle-node bifurcation, which can be reached for the case of larger optomechanical coupling than considered in this work.
%an unstable fixed point, which can be reached only in the case of strong optomechanical coupling.
%i.e., only if the radiation and/or thermal force are comparable to the elastic forces.
Note that the validity of the assumptions made in Sec.~\ref{sec:eq_of_motion}, especially the linear temperature dependence of the mechanical frequency and the thermal force, has to be carefully assessed in this case.

The Hopf bifurcation, which implies that periodic limit cycle oscillations can occur near the bifurcation threshold \cite{Guckenheimer&Holmes_book_83}, can readily be shown to correspond to a zero second Hurwitz determinant, i.e., $c_1c_2-c_3=0$, with a positive Hopf frequency $\omega_H=\sqrt{c_2}$. Using Eqs.~\eqref{eq:app_equilibrium_ci}, we find the bifurcation threshold condition to be
\begin{multline}
	\frac{\omega_0}{2Q}+\frac{\gamma_3}{2}A_{0s}^2+\eta\frac{\Omega_s}{\kappa^2+\Omega_s^2}\left(\beta A_{0s}+\frac{\theta}{2\Omega_s}\right)I'(A_{0s})\\
	=\frac{\frac{\omega_0}{2Q}+\frac{\gamma_3}{2}A_{0s}^2}{\kappa^2+\Omega_s^2} \Biggl[\nu I'(A_{0s})-3\alpha_3 A_{0s}^2\\
	-\kappa\left(\frac{\omega_0}{2Q}+\frac{\gamma_3}{2}A_{0s}^2\right)\Biggr].
	\label{eq:app_equilibrium_Hopf_cond}
\end{multline}
If we assume the mechanical dissipation, the nonlinear effects and the optomechanical coupling to be weak, namely, we assume the thermal frequency shift, the static displacement, the nonlinear and dissipation terms, the radiation pressure and the thermal force to be small, and, therefore, neglect all the small terms of the second order and higher, then the right hand side of Eq.~\eqref{eq:app_equilibrium_Hopf_cond} vanishes. In this limit, the Hopf bifurcation condition given in Eq.~\eqref{eq:app_equilibrium_Hopf_cond} coincides with the condition $\gamma=0$ discussed in Sec.~\ref{sec:small_osc_dynamics} [see Eqs.~\eqref{eq:small_osc_gamma} and \eqref{eq:small_osc_self_osc_threshold_cond}].

Under the same assumptions, the Hopf frequency becomes
\begin{equation}
	\omega_H=\sqrt{c_2}\approx \omega_0-\Delta\omega_s,
	\label{eq:app_equilibrium_wH}
\end{equation}
where $\Delta\omega_s$ is defined in Eqs.~\eqref{eq:small_osc_Dw0s} and \eqref{eq:small_osc_omegas}. This result coincides with the limit cycle frequency expression given in Eq.~\eqref{eq:small_osc_psi_lc} in the limit of vanishing limit cycle amplitude.

We note that the Hopf bifurcation can either be supercritical or subcritical, culminating with stable or unstable self-excited limit-cycle solutions which are discussed in Sec.~\ref{sec:self_osc}.

%------------------------------------------------------------------------------------------------------------------------------------

\section{Averaging of the equations of motion}
\label{app:averaging_eqs_motion}

Using Eqs.~\eqref{eq:large_osc_I_Fourier_kmax} and \eqref{eq:large_osc_x}, we write the optical power expression $I$ as
\begin{equation}
	I(x) \approx \sum_{k=-k_\text{max}}^{k_\text{max}} c_k e^{j2\pi \frac{k}{L}(A_0+A_1\cos\psi)}.
	\label{eq:app_large_osc_I_x_Fourier}
\end{equation}

It is beneficial to use the Jacobi-Anger expansion
\begin{equation}
	e^{jz\cos\xi}=J_0(z)+2\sum_{n=1}^{\infty}j^nJ_n(z)\cos n\xi,
\end{equation}
where $z$ and $\xi$ are some real variables, and $J_n(z)$ is the Bessel function of n-th order. The optical power expression given in Eq.~\eqref{eq:app_large_osc_I_x_Fourier} can be rewritten as
\begin{equation}
	I(x) \approx P_0+2\sum_{n=1}^\infty P_n\cos n\psi,
	\label{eq:app_large_osc_I_Bessel}
\end{equation}
where $P_n$ are defined in Eq.~\eqref{eq:large_osc_Pn}.

Next, we proceed to write the integral in Eq.~\eqref{eq:T-T0_orig} explicitly. Slow envelope approximation implies that the amplitude $A_1$, and the phase $\tilde\phi$ do not undergo significant changes at timescales comparable to  $\omega_0^{-1}$. It follows that $A_1$ and $\tilde\phi$ can be regarded as constants at timescales of order $\omega_0^{-1}$ and $\kappa^{-1}$, and terms involving $K$ in Eq.~\eqref{eq:full_eq_of_motion} can be estimated using the approximate equality
\begin{equation}
	\int_0^t f(\tau) g(\tau-t)d\tau \approx f(t)\int_0^t g(\tau-t) d\tau,
\end{equation}
where $g(\tau-t)$ is either $e^{\kappa(\tau-t)}$, $e^{(\kappa\pm j\omega_0)(\tau-t)}$ or $e^{(\kappa\pm j2\omega_0)(\tau-t)}$, $f(t)$ is a function of slow varying terms $A_1$ and $\tilde\phi$, and all fast decaying terms in $\int g(\tau-t)d\tau$ should be neglected. The result is
\begin{multline*}
	K(\cos n\psi)=\int_0^t \cos n\psi e^{\kappa(\tau-t)}d\tau \\
	\approx \frac{1}{2}\int_0^t \bigg( e^{(\kappa+jn\omega_0)(\tau-t)}e^{jn(\omega_0 t+\tilde\phi)}\\
	+e^{(\kappa-jn\omega_0)(\tau-t)}e^{-jn(\omega_0 t+\tilde\phi)} \bigg) d\tau \\
	=\frac{\kappa \cos n\psi+n\omega_0 \sin n\psi}{\kappa^2+n^2\omega_0^2},
\end{multline*}
\begin{equation}
	K(I)\approx \frac{P_0}{\kappa}+2\sum_{n=1}^{n_\text{max}} P_n \frac{\kappa\cos n\psi+n\omega_0\sin n\psi}{\kappa^2+n^2\omega_0^2}.
	\label{eq:app_large_osc_K_explicit}
\end{equation}

In order to solve Eq.~\eqref{eq:full_eq_of_motion} under the conditions described above, we use the harmonic balance method followed by the Krylov-Bogoliubov averaging technique \cite{Nayfeh_book_81}, and require
\begin{subequations}
\begin{align}
	x & =A_0+A_1\cos\psi, \nonumber \\
	\dot x & =-\omega_0 A_1\sin\psi,\\
	\ddot x & =-\omega_0^2 A_1\cos\psi-\omega_0\dot A_1\sin\psi-\omega_0 A_1\dot{\tilde\phi}\cos\psi.
\end{align}
	\label{eq:app_large_osc_x_deriv}
\end{subequations}
It follows that
\begin{equation}
	\dot A_1\cos\psi-A_1\dot {\tilde\phi}\sin\psi=0,
	\label{eq:app_large_osc_dxdt_condition}
\end{equation}

Introducing Eqs.~\eqref{eq:app_large_osc_x_deriv} into Eq.~\eqref{eq:full_eq_of_motion} results in
\begin{multline}
	-\omega_0^2A_1\cos\psi-\omega_0\dot A_1\sin\psi-\omega_0 A_1\dot{\tilde\phi}\cos\psi\\
	-\frac{\omega_0^2}{Q} A_1\sin\psi+\left[\omega_0-\beta\eta K(I)\right]^2(A_0+A_1\cos\psi)\\
	+\left[\alpha_3(A_0+A_1\cos\psi)-\gamma_3\Omega A_1\sin\psi\right](A_0+A_1\cos\psi)^2\\
	=2f_m\cos(\omega_0+\sigma_0)t+\nu I+\theta\eta K(I).
	\label{eq:app_large_osc_full_eq_of_motion}
\end{multline}

Collecting all non-harmonic terms in Eq.~\eqref{eq:app_large_osc_full_eq_of_motion} gives the expression for $A_0$:
\begin{multline}
	\alpha_3A_0^3+\left(\Omega^2+\frac32\alpha_3A_1^2 \right)A_0 \\
	=2P_1\beta\eta \frac{\omega_0\kappa}{\kappa^2+\omega_0^2}A_1+P_0\left(\nu+\frac{\theta\eta}{\kappa}\right),
	\label{eq:app_large_osc_A0_full}
\end{multline}
where $\Omega$ is defined in Eq.~\eqref{eq:large_osc_Omega},
% \begin{equation}
% 	\Omega=\omega_0-\frac{\beta\eta}{\kappa}P_0=\omega_0-\Delta\omega_0,
% 	\label{eq:app_large_osc_Omega}
% \end{equation}
and terms proportional to $(\beta\eta K)^2$ have been neglected because the frequency correction due to heating is considered small, i.e., $\beta\eta K(I) \ll \omega_0$. The term $\Delta\omega_0$ can be identified as a small frequency correction due to the heating of the mirror averaged over one mechanical oscillation period.

Equation~\eqref{eq:app_large_osc_A0_full} can be further simplified by assuming the static displacement $A_0$ to be small and using the weak nonlinearity assumption, i.e., $\alpha_3A_0^2\ll\omega_0^2$, giving rise to Eq.~\eqref{eq:large_osc_A0}.
% \begin{equation}
% 	A_0 \approx \frac{1}{\Omega^2+\frac32\alpha_3A_1^2}\left[2P_1\beta\eta \frac{\omega_0\kappa}{\kappa^2+\omega_0^2}A_1+P_0 \left(\nu+\frac{\theta\eta}{\kappa}\right)\right],
% 	\label{eq:app_large_osc_A0}
% \end{equation}

The remaining terms in Eq.~\eqref{eq:app_large_osc_full_eq_of_motion} constitute the following relationship [see also Eqs.~\eqref{eq:app_large_osc_I_Bessel} and \eqref{eq:app_large_osc_K_explicit}] :
\begin{equation}
	\dot A_1\sin\psi+A_1\dot{\tilde\phi}\cos\psi=B,
	\label{eq:app_large_osc_A1_and_phi_eq}
\end{equation}
where
\begin{multline}
	B=\Bigg[-\left(\frac{\omega_0}{Q}+\gamma_3 A_0^2+4P_2\beta\eta\frac{\omega_0}{\kappa^2+4\omega_0^2}\right)A_1\\
-\frac{\gamma_3}{4}A_1^3-2P_1\eta\frac{\omega_0}{\kappa^2+\omega_0^2}\left(2\beta A_0+\frac{\theta}{\omega_0}\right)\Bigg]\sin\psi \\
	+\Bigg[-\left(2\frac{\beta\eta}{\kappa}P_0-\frac{3\alpha_3}{\omega_0}A_0^2+2P_2\beta\eta \frac{\kappa}{\kappa^2+4\omega_0^2}\right)A_1+\frac{3\alpha_3}{4\omega_0}A_1^3\\
-2P_1\eta\frac{\kappa}{\kappa^2+\omega_0^2}\left(2\beta A_0+\frac{\theta}{\omega_0}\right)-2P_1\frac{\nu}{\omega_0}\Bigg]\cos\psi\\
-\frac{2f_m}{\omega_0}\cos(\omega_0+\sigma_0)t+\NST.
	\label{eq:app_large_osc_B}
\end{multline}
Here, $\NST$ denotes the non secular terms (i.e., higher harmonics).
% and
% \begin{equation}
% 	\Omega+\sigma_1=\omega_0+\sigma_0.
% 	\label{eq:large_osc_sigma1}
% \end{equation}

The Eqs.~\eqref{eq:app_large_osc_dxdt_condition} and \eqref{eq:app_large_osc_A1_and_phi_eq} can be rearranged as follows:
\begin{subequations}
\begin{align}
	\dot A_1 & =B\sin\psi,\\
	A_1\dot{\tilde\phi} & =B\cos\psi.
\end{align}
\label{eq:app_large_osc_full_B_eq}
\end{subequations}
Averaging of Eqs.~\eqref{eq:app_large_osc_full_B_eq} over one period of $\psi$ can be made under the assumption of slow varying envelope, namely:
\begin{subequations}
\begin{align}
	\dot A_1 & =\frac{1}{2\pi}\int_{0}^{2\pi}B\sin\psi d\psi,\\
	A_1\dot{\tilde\phi} & =\frac{1}{2\pi}\int_{0}^{2\pi}B\cos\psi d\psi.
\end{align}
\label{eq:app_large_osc_averaged_B_eq}
\end{subequations}

Substituting Eq.~\eqref{eq:app_large_osc_B} into Eqs.~\eqref{eq:app_large_osc_averaged_B_eq} yields the slow envelope evolution equations \eqref{eq:large_osc_polar_evolution_eq}.

%%%%%%%%%%%%%%%%%%%%%%%%%%%%%%%%%%%%%%%%%%%%%%%%%%%%%%%%%
\bibliographystyle{unsrtnat}
% \bibliography{c:/Technion/Research/Articles/GENERAL}
\bibliography{optomech_theory_arxiv.bbl}

\end{document}